\documentclass[%
 aip,
 amsmath,amssymb,
 reprint,%
]{revtex4-1}

\usepackage{graphicx}
\usepackage{dcolumn}
\usepackage{bm}
\usepackage{url}
\usepackage{hyperref}
\usepackage[utf8]{inputenc}
\usepackage[T1]{fontenc}
\usepackage{mathptmx}
\usepackage{etoolbox}\usepackage{comment}
\usepackage{bm}
\usepackage[dvipsnames]{xcolor}
\usepackage{graphicx}
\usepackage{dcolumn}
\usepackage{bm}
\newcommand{\highlight}[1]{%
  \colorbox{gray!20}{$\displaystyle#1$}}
\makeatletter
\def\@email#1#2{%
 \endgroup
 \patchcmd{\titleblock@produce}
  {\frontmatter@RRAPformat}
  {\frontmatter@RRAPformat{\produce@RRAP{*#1\href{mailto:#2}{#2}}}\frontmatter@RRAPformat}
  {}{}
}%
\makeatother
\begin{document}

\preprint{AIP/123-QED}


\title{Frequency combs induced by optical feedback and harmonic order tunability in quantum cascade lasers}
\author{Carlo Silvestri}
\affiliation{School of Electrical Engineering and Computer Science, The University of Queensland, Brisbane, QLD 4072, Australia
}

\author{Xiaoqiong Qi}%
\affiliation{School of Electrical Engineering and Computer Science, The University of Queensland, Brisbane, QLD 4072, Australia
}
\author{Thomas Taimre}
\affiliation{School of Mathematics and Physics, The University of Queensland, Brisbane, QLD 4072, Australia
}

\author{Aleksandar D. Raki\'c}
\affiliation{School of Electrical Engineering and Computer Science, The University of Queensland, Brisbane, QLD 4072, Australia
}

\date{\today}

\begin{abstract}
This study investigates the interaction between frequency combs and optical feedback effects in Quantum Cascade Lasers (QCLs). The theoretical analysis reveals new phenomena arising from the interplay between comb generation and feedback. By considering the bias current corresponding to free-running single mode emission, the introduction of optical feedback can trigger the generation of frequency combs, including both fundamental and harmonic combs. This presents opportunities to extend the comb region and generate harmonic frequency combs with different orders through optimization of external cavity parameters such as losses and length. Furthermore, the study demonstrates that optical feedback can selectively tune the harmonic order of a pre-existing free-running comb by adjusting the external cavity length, particularly for feedback ratios around $1\%$, which are readily achievable in experimental setups. Under strong feedback conditions (Acket parameter $C>4.6$), mixed states emerge, displaying the features of both laser and external cavity dynamics. While the study is predominantly centered on Terahertz QCLs, we have also confirmed that the described phenomena occur when utilizing mid-infrared QCL parameters. This work establishes a connection between comb technology and the utilization of optical feedback, providing new avenues for exploration and advancement in the field. In fact, the novel reported phenomena  open a pathway towards new methodologies across various domains, such as design of tunable comb sources, hyperspectral imaging, multimode coherent sensing, and multi-channel communication.\footnote{The following article has been accepted by APL Photonics. After it is published, it will be found at https://pubs.aip.org/aip/app.}

\end{abstract}

\maketitle

\section{Introduction}
\begin{figure*}[t]
\centering
\includegraphics[width=0.83\textwidth]{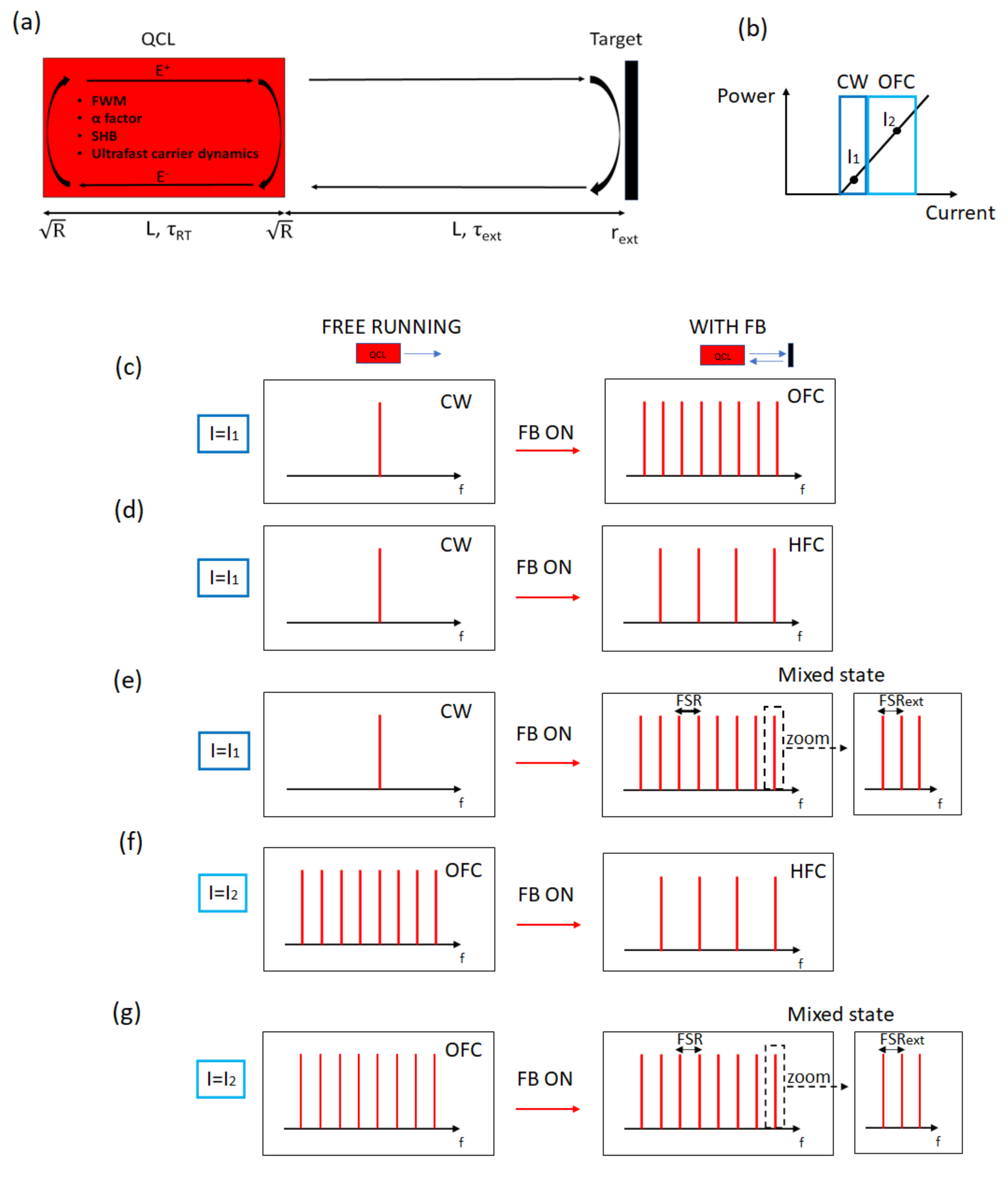}
\caption{(a) QCL in presence of an external target. The QCL cavity has length $L$ and roundtrip time $\tau_\mathrm{RT}$, and it is delimited by facets with reflectivity for the electric field $\sqrt{R}$, where $R$ is the power reflectivity. The reflectivity of the target is $r_\mathrm{ext}$, the EC has length $L_{\mathrm{ext}}$ and roundtrip time $\tau_{\mathrm{ext}}$. (b) Schematic representation of the Power-Current curve for the free-running QCL, with highlighted the CW emission region close to threshold (light blue), and the comb region named OFC (dark blue); $I_\mathrm{1}$  and $I_\mathrm{2}$ are values of the bias current respectively in the CW and OFC region. The details on the simulated free-running QCL can be found in the supplementary material and in ref.\cite{Silvestri22}. (c)-(g): frequency domain schematic of the novel reported phenomena induced by the optical feedback; feedback induced transition from a free-running CW ($I=I_\mathrm{1})$ to OFC (c), HFC (d), and mixed state (e), and from free-running OFC ($I=I_\mathrm{2})$ to HFC (f), and mixed state (g). $\mathrm{FSR}$: free spectral range of the QCL cavity. $\mathrm{FSR_\mathrm{ext}}$: free spectral range of the external cavity.}
\label{Figure1}
\end{figure*}

Quantum cascade lasers (QCLs) are unipolar semiconductor lasers based on electronic transitions between confined states in the conduction band, named subbands.\cite{Faist_1994} The QCL active region is a multi-stage heterostructure made of nanometric semiconductor layers, whose thickness determines the energy associated to the electronic transitions.\cite{Faist_book,Vitiellorev} For this reason, these devices are characterized by bandgap tunability, providing emission in the terahertz (THz) \cite{Kohler2002} and mid-infrared (mid-IR) \cite{Faist_1994} spectral regions.\\
In 2012, it was demonstrated for the first time that QCLs can spontaneously generate optical frequency combs (OFCs),\cite{Hugi2012} coherent dynamical regimes consisting of multiple phase-locked optical lines.\cite{Hugi2012,Faist_2016,Burghoff2014} OFCs arise in QCLs without any external component such as radiofrequency injection, or passive mode-locking, because of the interplay between strong nonlinearities, non-zero linewidth enhancement factor (LEF, also known as $\alpha$ factor), spatial-hole burning (SHB), and ultrafast carrier lifetime.\cite{SilvestriReview,PiccardoReview} Over the past decade the research about QCL combs has been very fertile, producing the discovery and characterization of several different types of these states, such as harmonic frequency combs (HFCs), combs induced by phase turbulence, frequency modulated OFCs, and solitons.\cite{Hugi2012,Burghoff2014, Opacak2019,PiccardoReview,Faist_2016, Singleton18,Unifying,Pratichaos,Pratinano,NaturePiccardo,Bomeng2} Strategies to optimize and manipulate these coherent states have been proposed in the literature, exploiting RF injection and external optical field injection, and resulting in improvements in terms of stability, spectral broadening, and the ability to induce harmonic states.\cite{schneider, consolino1, PiccardoOptical}\\
Another convenient possibility for modifying and optimizing QCL combs is represented by the use of optical feedback. This approach simplifies the experimental setup and offers potential solutions for coherent sensing, spectroscopy and communication. Previous studies on QCLs under optical feedback have primarily focused on single-mode operation,\cite{Rakicreview,mezzapesa2013, Taimre15} which exhibits higher stability against feedback compared to conventional laser diodes (LDs).\cite{mezzapesa2013,Grillot14} This arises from the combination of two peculiar properties of these lasers: ultrafast carrier lifetime, which leads to a suppression of the relaxation oscillations, and a LEF lower than in interband LDs, which corresponds to a reduced coupling between amplitude and phase of the electric field.\cite{mezzapesa2013,Grillot16} Bandgap tunability and stability under feedback have lead to the employment of single mode distributed-feedback (DFB) QCLs for several applications in the sensing domain particularly in the THz region, where these lasers supply a convenient low-background detection system and offer a solution to the lack of efficient detectors.\cite{Rakicreview,Pogna2021, Giordano18, Lim19,SilvestriSNOM}\\
Conversely, the study of the effects of optical feedback on QCL combs is still in its early stages, with a limited number of published studies documented in the literature. These initial works, all of an experimental nature, have shown that the presence of an external target contributes to improving the stability of the combs, with the observation of a narrowing of the intermode beatnote,\cite{VitielloSNOM2022, Teng19} and also allows for slight variation (about $1\%$) in the mode spacing of these coherent states.\cite{Teng23} Furthermore, by fine-tuning the laser-target distance on the emission wavelength scale, a periodic evolution of the beatnote from single to multi-peak was observed, with period equal to half the wavelength.\cite{Liao22} In addition, we mention that defects and reflectors have been utilized to control the combs emitted by QCLs in some studies.\cite{Kazakov21,Hillbrand18}\\
However, no systematic theoretical investigation on the combination of feedback and QCL combs has been conducted so far. This work fills this gap in the existing literature and provides a comprehensive study of the feedback effects in QCL OFCs by utilizing a new theoretical approach. In fact, the laser dynamics in presence of feedback has been generally investigated by employing Lang--Kobayashi (LK) \cite{LangKobayashi} or reduced rate equations,\cite{Tina1,Tina2} which do not include the polarization dynamics and dispersive effects necessary to reproduce OFCs. Our study includes the phenomena responsible for comb formation and accounts for the interplay between laser and external cavity (EC) modes in the QCL medium in the presence of a back-reflecting target (Fig.~\ref{Figure1}(a)). For this purpose, we adopt a full set of effective semiconductor Maxwell--Bloch equations (ESMBEs) for a Fabry--Perot (FP) QCL,\cite{Silvestri20, SilvestriCLEO,Silvestri22} modified by incorporating the optical feedback into the model.\\ 
It is worth noting that the ESMBEs allow for reproducing dynamical scenarios by varying the bias current in accordance with experiments on free-running QCLs,\cite{Li15,SilvestriReview} showing a near-threshold continuous-wave (CW) emission followed by a comb region (Fig.~\ref{Figure1}(b)). Therefore, by selecting the driving current of the QCL, we can investigate how different free-running regimes (CW and OFC) evolve with the introduction of the optical feedback. This leads to the prediction of several new phenomena, graphically represented in Fig.~\ref{Figure1}(c)-(g), and summarized below:
\begin{enumerate}
\item The optical feedback can destabilize the single mode emission and trigger the generation of fundamental (Fig.~\ref{Figure1}(c)) and harmonic (Fig.~\ref{Figure1}(d)) combs; HFCs with different mode spacing are obtained by varying EC length and feedback strength.
\item Under strong feedback (Acket parameter $C>4.6$) the feedback induces a transition from single mode to mixed states displaying the features of both, laser and external cavity (Fig.~\ref{Figure1}(e)). In the frequency domain, mixed states correspond to a set of frequency bands whose central frequencies are the modes of the laser cavity, and whose spacing is the free spectral range of the EC $\mathrm{FSR_{ext}}$.
\item A transition from a fundamental to a harmonic comb can be induced by introducing optical feedback (Fig.~\ref{Figure1}(f)).
\item Under strong feedback, a transition from a fundamental comb to a mixed state can be triggered by optical feedback (Fig.~\ref{Figure1}(g)).
\end{enumerate}
These findings, in addition to being relevant from a fundamental physics perspective, hold great significance for the advancement of tunable comb sources, multimode coherent sensing, multi-channel communication, and the extension of the dynamical range of QCL combs.\\
Section~\ref{model} is dedicated to describing the main features of the ESMBEs and illustrates how the boundary conditions are modified to incorporate optical feedback into the model.\\
In Section~\ref{SecCW}, we present and analyze the results obtained in the presence of feedback when the bias current of the laser is set to a value corresponding to single-mode emission in free-running operation. We also discuss the influence of the $\alpha$ factor and carrier lifetime on the dynamical scenario.\\
Section~\ref{SecOFC} is about the dynamics of the QCL in the presence of feedback when the bias current corresponds to the emission of a pre-existing comb in free-running operation.\\
Section~\ref{SecConcl} summarizes the conclusions of the study and highlights the potential applications of the theoretically predicted novel phenomena.
\section{The theoretical model}\label{model}
The ESMBEs encompass the polarization dynamics and the main properties of semiconductor materials playing a role in the comb formation, such as non null $\alpha$ factor, dependence of the optical susceptibility from the carrier density, four-wave mixing, and SHB.\cite{SilvestriReview} Two counterpropagating forward and backward fields, respectively $E^+$ and $E^-$, are considered.\\
In this work we introduce the optical feedback in the ESMBEs by modifying the FP boundary conditions with respect to the free running operation.\cite{Rimoldi22} For this purpose, we consider the case of an external target placed at distance $L_{\mathrm{ext}}$ from the right facet of the QCL, with a frequency independent reflectivity $r_\mathrm{ext}$, $\epsilon_\mathrm{L}$ quantifies the total losses experienced from the field in the EC, and $\epsilon_\mathrm{S}$ the ones due to the re-injection. The field emitted by the QCL travels in the EC, is reflected by the target, and partially re-enters into the laser cavity (Fig.~\ref{Figure1}(a)). If $\mathrm{\epsilon}=r_\mathrm{ext}\epsilon_\mathrm{L}\epsilon_\mathrm{S}$, we can write the boundary conditions as follows:\\
\begin{eqnarray}
E^-(L, t)&=&\sqrt{R}E^+(L,t)\nonumber\\&+&\mathop{\highlight{\mathrm{\epsilon} t_L^2 E^+(L, t-\tau_\mathrm{ext})\exp(-i\omega_0\tau_\mathrm{ext})}}^{\mathrm{optical~feedback}},\label{bc1fb}\\
E^+(0, t)&=&\sqrt{R}E^-(0,t),\label{bc2fb}
\end{eqnarray}\\
where $L$ is the length of the QCL cavity, $R$ is the power reflectivity of both the QCL facets, $t_L=\sqrt{1-R}$ is the transmissivity of the laser facet for both the outgoing and the incoming field, $\omega_0$ is the central value of the angular frequency, and $\tau_{\mathrm{ext}}=\frac{2L_{\mathrm{ext}}}{c}$ is the roundtrip time in the EC. Therefore the feedback is included by adding the highlighted term at the right hand side in Eq. (\ref{bc1fb}). The ratio between the feedback power and the output power is $\epsilon^2$. For $\epsilon=0$ the free-running case is obtained. A full description of the ESMBEs and the mathematical procedure exploited to include the optical feedback, are presented in the supplementary materials.
\section{Feedback effects starting from single-mode emission}\label{SecCW}
We study the dynamics of a QCL in the presence of an EC by integrating the ESMBEs with the boundary conditions (\ref{bc1fb})--(\ref{bc2fb}). The duration of each simulation is around $1-5~\mu\mathrm{s}$, ensuring that the system reaches a steady-state condition. In this section we consider the bias current of the laser set to $I=1.08I_\mathrm{thr}$ ($I_\mathrm{thr}$ is the threshold current), which corresponds to a case of single mode emission in free-running operation, and we investigate the effect of the feedback. 
\begin{figure*}[t]
\centering
\includegraphics[width=1\textwidth]{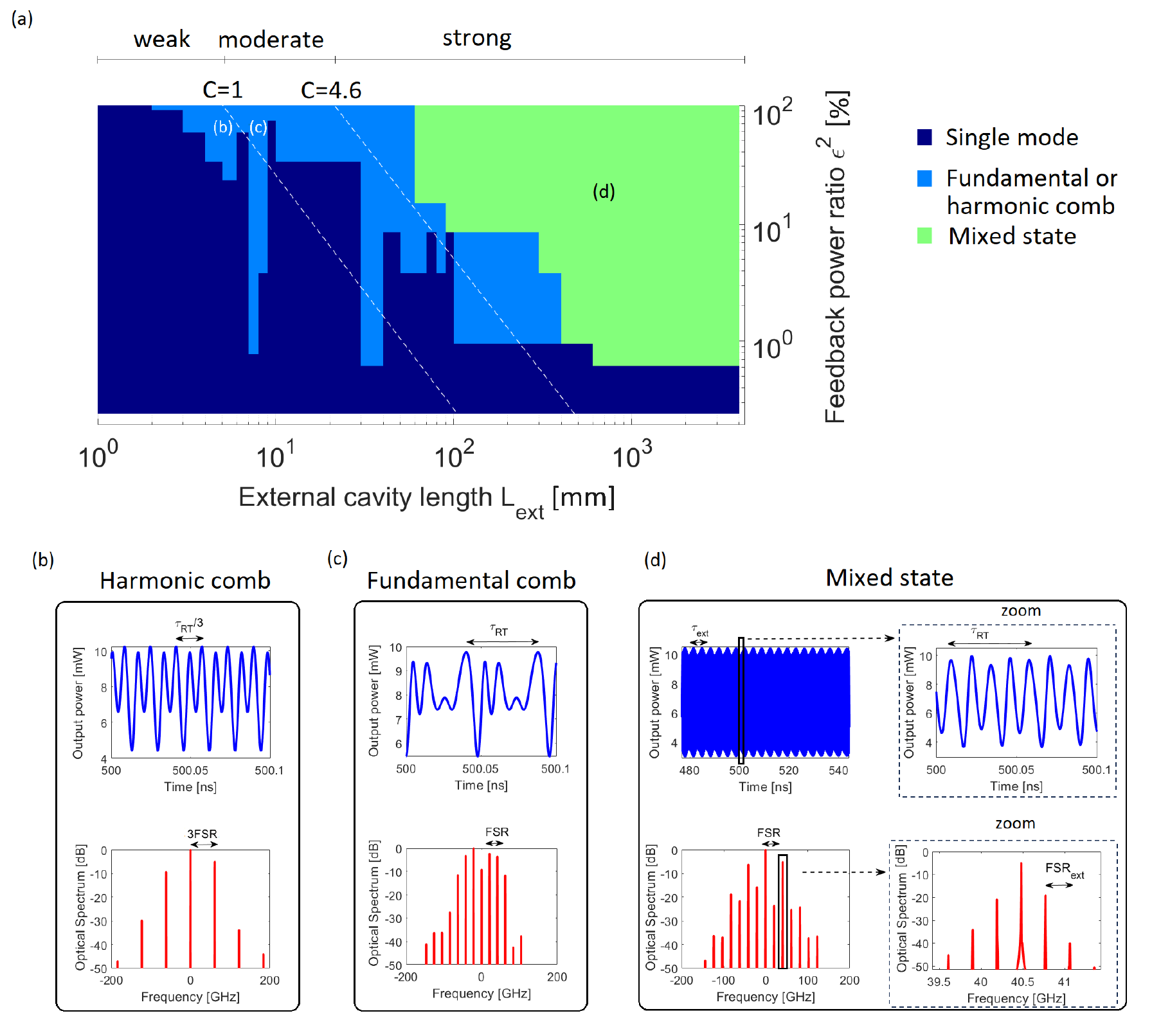}
\caption{(a) Feedback regime map for a Fabry--Perot THz QCL in the space of EC length $L_\mathrm{ext}$ (horizontal axis) and feedback power ratio $\epsilon^2$ (vertical axis). The bias current is $I=I_\mathrm{1}=1.08I_\mathrm{thr}$, which corresponds to CW emission in free-running operation. 600 simulations have been performed, by varying $\epsilon$ between 0.01 and 0.09 with step 0.01 and between 0.1 and 1 with step 0.1, while the values of $L_\mathrm{ext}$ are considered in four regions: 1~mm $\leq L_\mathrm{ext} \leq$ 9~mm, with step 1~mm; 1~cm $\leq L_\mathrm{ext}\leq$ 9~cm, with step 1~cm; 10~cm $\leq L_\mathrm{ext}\leq$ 90~cm, with step 10~cm; 1~m $\leq L_\mathrm{ext}\leq$ 3~m, with step 1~m. The values of $L_\mathrm{ext}$ and $\epsilon^2$ are displayed in logarithmic scale. The region $\epsilon^2<0.25~\%$ is characterized by CW emission for every pair ($L_\mathrm{ext}$, $\epsilon^2$) and it is not shown. The white dashed lines represent the points of the plane where the Acket parameter $C$ has values 1 and 4.6, as indicated. Examples of found dynamical regimes are shown in the panels (b)--(d) and highlighted on the map of Fig. (a): output power (up) and optical spectrum (down) of third order harmonic (b) and fundamental (c) combs, mixed state.}
\label{Figure2}
\end{figure*}
The results for a THz QCL with carrier lifetime $\tau_\mathrm{e}=5~$ps,\cite{SilvestriReview, PiccardoReview, THz_taue_1} and linewidth enhancement factor $\alpha=-0.1$ \cite{alphaneg, Silvestri22, Tina1, negalpha1,negalpha2} are summarized in the map presented in Fig.~\ref{Figure2}(a). In the dark blue region the free-running CW emission is unchanged by the introduction of the feedback. This behaviour tends to disappear as the feedback level increases ($\epsilon^2>1\%$). For values of $L_\mathrm{ext}$ between $2~$mm and $10~$cm (short external cavity regime) the optical feedback destabilizes the single mode emission and triggers the formation of comb regimes (light blue region), both fundamental (Fig.~\ref{Figure2}(c)) and harmonic (Fig.~\ref{Figure2}(b)).  The majority of these states occur for $\epsilon^2>10\%$, when the feedback power constitutes a relevant portion of the output power. However, if $L_\mathrm{ext}$ is between $10~$cm and $40~$cm, some OFC regimes are found for feedback ratio around $1\%$, which is a condition that is easier to achieve in experiments. We remark that in the short cavity range, $\mathrm{FSR}_{\mathrm{ext}}$ has the same order of magnitude of the laser cavity $\mathrm{FSR}$, with important implications for the generation of states with different harmonic order. This aspect will be analyzed in detail in the second part of this section.\\
If we leave the condition of short cavity, $\mathrm{FSR}_{\mathrm{ext}}$ and $\mathrm{FSR}$ have different order of magnitude, and the features of the both the cavities are identifiable in the QCL dynamics. In fact, the dynamical scenario is dominated by a different class of regimes (green region in  Fig.~\ref{Figure2}(a)) for $L_\mathrm{ext}>10~$cm, and for $\epsilon^2>1\%$. An example of this type of emission is shown in Fig.~\ref{Figure2}(d). The temporal evolution of the output power is characterized by modulations with periodicity given by the roundtrip time of the EC $\tau_\mathrm{ext}$ (up-left panel in Fig.~\ref{Figure2}(d)), but it simultaneously presents oscillations on the time scale of the QCL cavity roundtrip (zoom of the output power, up-right panel in Fig.~\ref{Figure2}(d)). Therefore, we can define these regimes as mixed states, since they display the features of both laser and external cavity. Consequentially, by looking at the optical spectrum, we observe that each fundamental mode of the laser cavity presents a fine structure composed by a few secondary peaks around the main one (down panels in Fig.~\ref{Figure2}(d)). We have, therefore, a sequence of frequency bands, each of them centered at one of the longitudinal modes of the laser cavity, and with spacing corresponding to $\mathrm{FSR}_\mathrm{ext}$. These regimes can result particularly promising for application purposes, since the spacing of each frequency band is tunable with continuity by varying $L_\mathrm{ext}$, and at the same time the central frequency can be chosen between the fundamental longitudinal modes of the QCL cavity. Indeed, they could be useful for hyperspectral imaging in order to measure the spectral fingerprints of a material within the frequency range covered by the mixed states. Furthermore, they could be appealing for the implementation of novel multi-channel communication systems.\\
It is worth to notice that the scenario described in Fig.~\ref{Figure2} presents a correspondence with the classical feedback regimes defined by the Acket parameter $C= \frac{\epsilon (1-R)\tau_\mathrm{ext}}{\sqrt{R}\tau_\mathrm{RT}}\sqrt{1+\alpha^2}$.\cite{Taimre15} In Fig.~\ref{Figure2}(a) the different regions of feedback regimes defined by $C$ are delimited by white dashed lines, and we can notice that single mode emission and feedback-induced combs correspond approximately to the union of weak and moderate feedback regions, while the mixed states arise under strong feedback ($C>4.6$). We specify that for $\epsilon^2<0.25\%$ we found stable single mode emission for each value of $L_\mathrm{ext}$, with negligible effect of the feedback. A similar behaviour (stability of the CW emission for a feedback ratio lower than approximately 0.1$\%$) was reported also in the feedback diagram for single-mode DFB QCLs presented in ref.~\cite{Grillot14}. It is a manifestation of the higher stability of QCLs mentioned in the introduction, due to lower $\alpha$ and carrier lifetime with respect to bipolar LDs. Lastly, we highlight that the classification of the different dynamical regimes presented in the map of Fig.~\ref{Figure2} has been performed using a rigorous procedure based on phase and amplitude noise quantifiers previously introduced in ref.~\cite{Silvestri20}. The details on this procedure can be found in the supplementary materials.\\\\
\begin{figure}[t]
\centering
\includegraphics[width=0.5\textwidth]{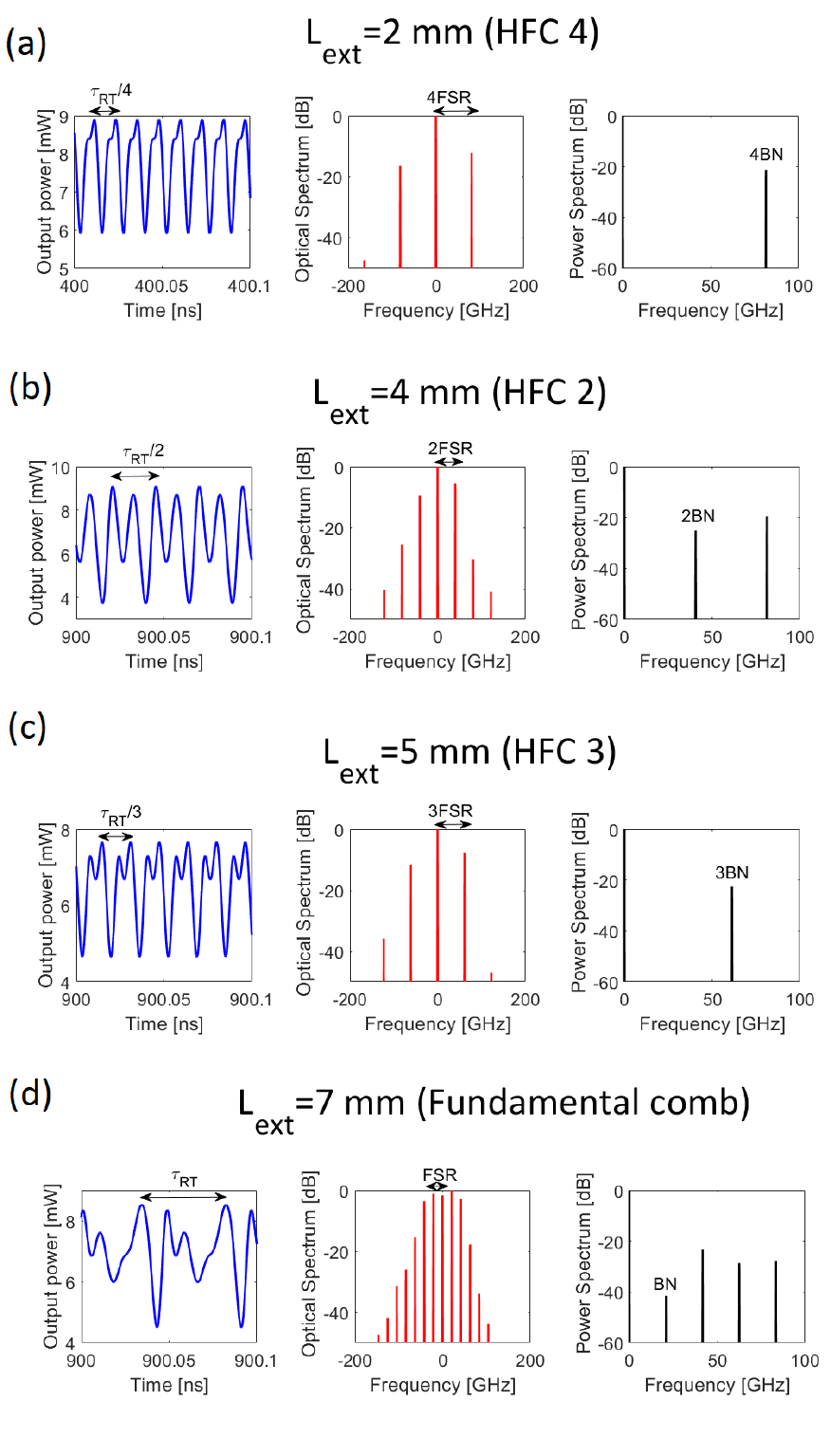}
\caption{Harmonic order tunability in the short cavity range. Feedback-induced frequency combs with different harmonic order obtained for different values of $L_\mathrm{ext}$: (a) fourth, (b) second, (c) third order HFCs, and (d) fundamental optical frequency comb, reported respectively for $L_\mathrm{ext}=2$~mm, $4~$mm, $5~$mm and $7~$mm. The combs of Figs.~(b)-(d) are obtained for the same feedback ratio $\epsilon^2=36~\%$, while the HFC 4 of Fig.~(a) is found for $\epsilon^2=100~\%$. Temporal evolution of the output power (left), optical spectrum (centre), and power spectrum (right) are shown for each regime. These regimes are extracted from Fig.~\ref{Figure2}(a), so they are generated for $I=1.08I_\mathrm{thr}$, corresponding to a CW emission in free-running operation.}
\label{Figure3}
\end{figure}
At this point, we want to discuss more in detail the formation of the comb regimes presented in the map of Fig.~\ref{Figure2}(a). Firstly, we highlight that optical feedback induces new types of combs which are not observed in free running for any value of the bias current, by adopting the exploited set of parameters. In fact, in that case only self-starting fundamental and second order HFCs are reported,\cite{Silvestri22} while in presence of optical feedback we observe third and fourth order HFCs, as shown in Fig.~\ref{Figure3}. In particular, the short external cavity regime offers the possibility to tune the order of the generated frequency combs by varying the external cavity length $L_\mathrm{ext}$. In Fig.~\ref{Figure3}(b)-(d) we show some examples of HFCs characterized by different order, obtained for different values of $L_\mathrm{ext}$, at fixed $\epsilon^2=36\%$. We specify that these regimes are elements of the map presented in Fig.~\ref{Figure2}(a) in the main text, so that they are generated for $I=1.08I_\mathrm{thr}$, corresponding to a single mode emission in free-running operation. Therefore, these regimes are induced by the feedback. Moreover, we observe that the order of these combs is related to the chosen values of $L_\mathrm{ext}$. For the case in Fig.~\ref{Figure3}(b) we have $L_\mathrm{ext}=4~$mm, which corresponds to $\mathrm{FSR}_\mathrm{ext}=37.5~$GHz. Considering that the QCL cavity $\mathrm{FSR}$ is $20.8~$GHz, we observe that in this case $\mathrm{FSR}_\mathrm{ext}$ is close to $2\mathrm{FSR}$, and in fact the generated comb has order 2. If $L_\mathrm{ext}=5~$mm we report a third order HFC regime (Fig.~\ref{Figure3}(c)), and we have $\mathrm{FSR}_\mathrm{ext}=30~$GHz$\approx1.5~\mathrm{FSR}$, so that the lowest integer which is a multiple of $\frac{\mathrm{FSR}_\mathrm{ext}}{\mathrm{FSR}}$ is 3 (this corresponds to a superposition between all the external cavity modes with the modes of the cold QCL cavity spaced by 3~$\mathrm{FSR}$). When $L_\mathrm{ext}=7~$mm, we have $\mathrm{FSR}_\mathrm{ext}=21.4~$GHz, a value close to the $\mathrm{FSR}$ and in fact, as shown in Fig.~\ref{Figure3}(d), the feedback induces a fundamental OFC. We also verified that, by choosing the values of $\L_\mathrm{ext}$ so that the conditions $\mathrm{FSR}_\mathrm{ext}=\mathrm{FSR}$, $\mathrm{FSR}_\mathrm{ext}=1.5~\mathrm{FSR}$, and $\mathrm{FSR}_\mathrm{ext}=2\mathrm{FSR}$ are strictly satisfied, we obtain respectively combs of order 1, 3, and 2 as in Fig.~\ref{Figure3}. The results presented in Fig.~\ref{Figure3} help us to understand that the order of these combs is determined by $\mathrm{FSR}_\mathrm{ext}$, so that the external cavity modes compete in the nonlinear QCL medium and impose their characteristics, generating a comb with their spacing (Figs.~\ref{Figure3}(b)-\ref{Figure3}(d)) or with a spacing which is an integer multiple of $\mathrm{FSR}_\mathrm{ext}$ (Fig.~\ref{Figure3}(c)). Finally, we observe that the feedback strength plays a role in the possibility to generate harmonic combs with different order, and a larger variety of these regimes is reported as $\epsilon^2$ increases. For example, if we increase $\epsilon^2$ to $100\%$ (no losses in the external cavity), we also report a 4th order HFC (see Fig.~\ref{Figure3}(a)) when $L_\mathrm{ext}=2~$mm, corresponding to $\mathrm{FSR}_\mathrm{ext}=75~$GHz (close to $4~\mathrm{FSR}$). We would like to clarify that while achieving a feedback ratio of $100\%$ may not be feasible in practice, this example nonetheless highlights the possibility of 4th order HFC and the sequential switching of comb regimes with increasing feedback levels.\\
\\We understand that the addition of an external target implies a three mirror cavity (as depicted in Fig.~\ref{Figure1}(a)) with a modified spectral behaviour with respect to the two mirror laser cavity in free running operation, so that the emission of new comb states is triggered.\\
According to these results, optical feedback can serve as a new method to tune the harmonic order of QCL combs, which is also more convenient and cost-effective compared to methods based on optical injection.\cite{PiccardoOptical} This could enable significant improvements in the field of broadband spectroscopy, imaging, and wireless communication, which are the most compelling applications of QCL HFCs.\cite{PiccardoHFCOptex}\\
Additionally, these results show that optical feedback helps the OFC formation closer to the laser threshold, with a consequent extension of the comb region in terms of bias current, overcoming the limitations imposed by the typically short dynamical range of QCLs. Although the short-cavity region requires high feedback ratios (>10\%) to trigger the combs, for $L_\mathrm{ext}$ between 10 and 40 cm, these regimes are induced for $\epsilon^2$ on the order of 1\%, as mentioned previously. Therefore, we identify this portion of the feedback map in Fig.~\ref{Figure2}(a) as ideal for applications related to extending the comb region.\\
In addition to the results shown in Fig.~\ref{Figure2}, we have also investigated how the comb emission is affected when fine-tuning the length of the external cavity on the wavelength scale. In particular, we observed that for the case $I=1.08I_\mathrm{thr}$, the comb emission is obtained with a period of $\lambda$/2 ($\lambda$ is the central emission wavelength) and alternates with single-mode emission in the short cavity regime. Conversely, when shifting $\lambda$ on the wavelength scale, we consistently obtain comb emission in the long cavity regime, although we report variations in the waveform. Details regarding these results are presented in section S.2 of the supplementary materials for the case $I=1.08I_\mathrm{thr}$.

\begin{figure}[t] 
\centering
\includegraphics[width=0.5\textwidth]{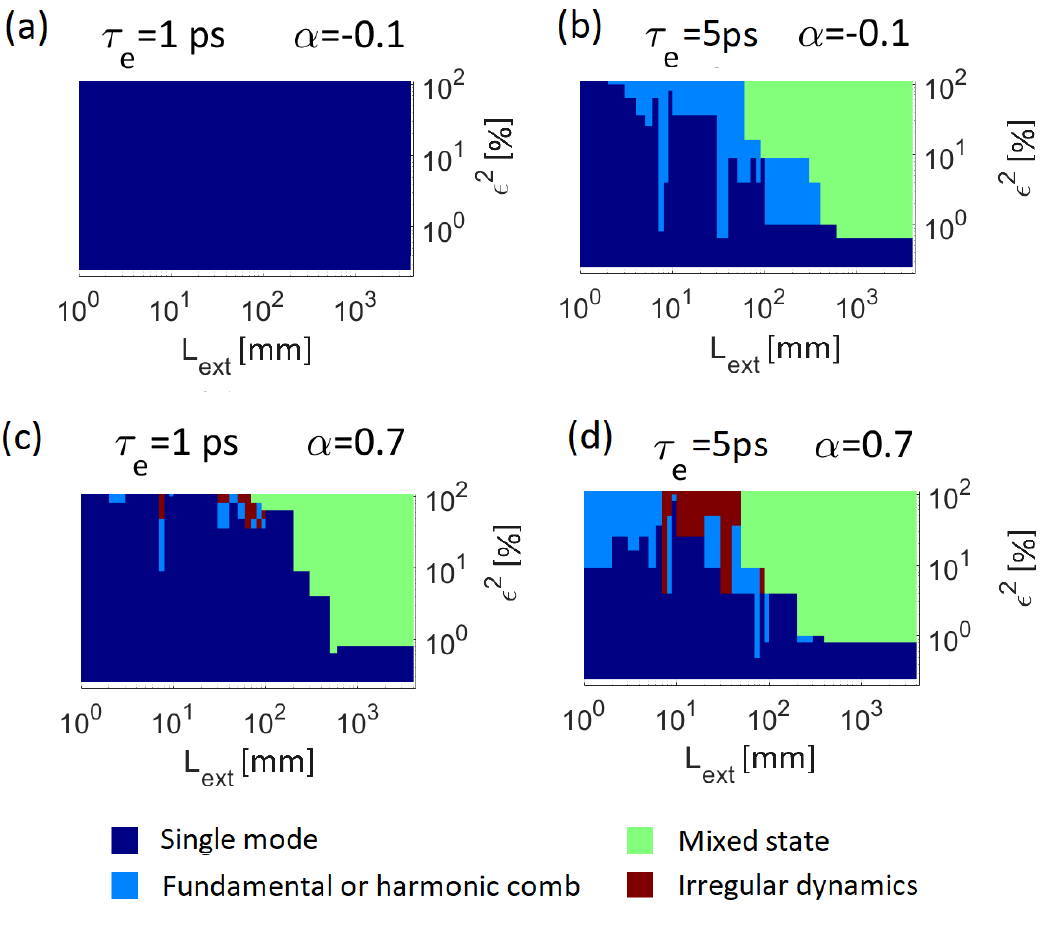}
\caption{Feedback maps showing the dynamical regimes for different values of  $\tau_\mathrm{e}$ and $\alpha$.  Values of $\epsilon$ and $L_\mathrm{ext}$ as in Fig.~\ref{Figure2}(a). $I=1.08I_\mathrm{thr}$, and other parameters as in Fig.~\ref{Figure2}(a).}
\label{Figure4}
\end{figure}

Then, we investigate the impact of carrier lifetime and $\alpha$ factor on the feedback regime scenario, considering the four combinations between two values of $\tau_\mathrm{e}$ (1~ps and 5~ps) and two values of $\alpha$ ($-0.1$ and 0.7). We remark that $\tau_\mathrm{e}=$ 1~ps and $\alpha=0.7$ are typical values for a mid-IR QCL,\cite{SilvestriReview,PiccardoReview} while $\tau_\mathrm{e}=$ 5~ps and $\alpha=-0.1$ are characteristic of THz-QCLs, as previously mentioned. This last combination, in fact, has been previously analyzed and described in Fig.~\ref{Figure2}, and the map of Fig.~\ref{Figure2}(a) is again reproduced in Fig.~\ref{Figure4}(b), to enable comparison with the other cases.
We can notice that if we keep $\alpha=-0.1$, but we decrease the value of the carrier lifetime to 1~ps, the CW emission dominates the dynamical scenario in Fig.~\ref{Figure4}(a), and the comb and mixed state regions disappear. This suggests that the low value of $\tau_\mathrm{e}$ provides ultrastability of the single mode-emission, in agreement with previous studies on QCLs under optical feedback based on the LK model, where it was shown that an increase of the photon to carrier lifetime ratio implies higher stability of the single mode solution.\cite{mezzapesa2013,columbo2014} We would like to point out that we have recreated the feedback diagram for two distinct values of $\tau_\mathrm{e}$ in order to theoretically investigate how the laser dynamics are influenced by this parameter. Our aim was not to reproduce a variation in carrier lifetime that can be observed in a laboratory setting for a single device.\\If we increase $\alpha$ to 0.7 by keeping $\tau_\mathrm{e}=1$~ps (Fig.~\ref{Figure4}(c)), islands of comb regimes and mixed states reappear in the map, and we also observe a new type of states, characterized by irregular dynamics (red region), which do not present locking and do not display any periodicity on the EC roundtrip time. We relate this to the higher value of $\alpha$, associated to a higher phase-amplitude coupling of the electric field, which favors the occurrence of multi-mode regimes and can lead to chaotic or irregular dynamics when it is high enough, as shown for free-running QCLs.\cite{Silvestri20} In this case the comb region is less extended than in the THz QCL case of Fig.~\ref{Figure4}(b), and only few locked states are reported mainly for high feedback coupling ($\epsilon^2>10\%$). We explain this by considering that the low value of $\tau_\mathrm{e}$ tends to keep the system on a stable CW emission, as observed in the limit case of Fig.~\ref{Figure4}(a). We remark that even if the values $\alpha=0.7$ and $\tau_\mathrm{e}=1~$ps are typical of mid-IR QCLs, \cite{SilvestriReview,PiccardoReview} the other parameters used to generate Fig.~\ref{Figure4}(c) are common to Fig.~\ref{Figure2}(a) and to all the other maps in Fig.~\ref{Figure4}, and correspond to a THz QCL. A feedback map for an actual mid-IR QCL is presented and discussed in the supplementary materials. In this case a comb region more extended than in Fig.~\ref{Figure4}(c) is found, and it is shown that this is linked to the larger value of gain bandwidth characterizing the mid-IR devices.\cite{Faist_2016,Singleton18} This gives generality to our results, because it assures that both THz and mid-IR QCLs can provide feedback-induced comb operation in a large portion of the $L_\mathrm{ext}$--$\epsilon^2$ diagram.\\
Lastly, for $\alpha=0.7$ and $\tau_\mathrm{e}=5~$ps, we observe a larger number of multi-mode regimes, both locked and unlocked (Fig.~\ref{Figure4}(d)). We understand, therefore, that an increase of the carrier lifetime promotes a destabilization of the CW emission when the feedback is switched on, playing a role similar to the LEF. However, we highlight that an increase of $\alpha$ implies the emergence of a higher number of irregular regimes. We estimated a critical value $\alpha=0.5$ for which irregular dynamics arises (see the supplementary materials). This value is compatible with THz QCLs,\cite{SilvestriReview} suggesting the possibility to generate irregular or chaotic regimes in these devices in presence of feedback.
\section{Feedback effects starting from comb emission}\label{SecOFC}
\begin{figure}[t]
\centering
\includegraphics[width=0.5\textwidth]{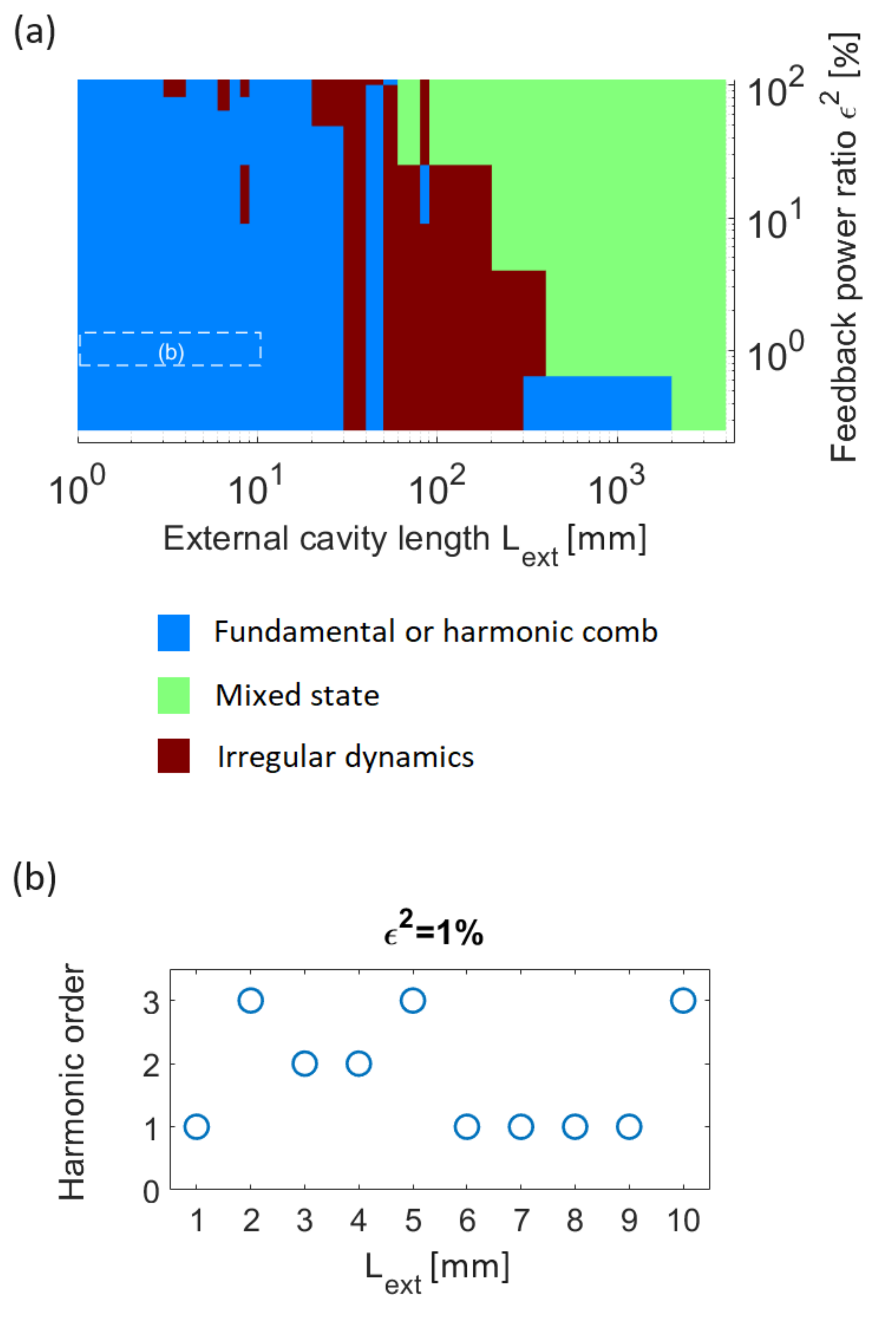}
\caption{(a) Feedback regime diagram for $I=I_2=1.5I_\mathrm{thr}$, corresponding to the emission of a fundamental OFC in free-running operation. All the other parameters as in Fig.~\ref{Figure2}(a). (b) Order of frequency combs reported for different values of the external cavity length in regime of short cavity, for feedback ratio $\epsilon^2=1\%$. The data are extracted from the map of Fig.~(a), where the region considered in the panel (b) is highlighted with a dashed box.}
\label{Figure5}
\end{figure} 
We investigate how the map of Fig.~\ref{Figure2}(a) changes if we vary the bias current,  using a frequency comb as the initial condition in our numerical experiment instead of a CW emission. We realize this by setting $I=1.5I_\mathrm{thr}$, where the QCL emits a fundamental frequency comb regime in free-running operation, and we replicate the study of  Fig.~\ref{Figure2} in this new case. The resulting map, shown in Fig.~\ref{Figure5}(a), exhibits three types of regimes: frequency combs, occurring mainly for short values of the cavity length, mixed states, corresponding to $L_\mathrm{ext}>10~$cm, and irregular dynamics. We observe, therefore, that the single-mode region disappears if we choose a comb as initial condition. This indicates that the dynamical scenario exhibited from the QCL depends on the bias current, and different maps are obtained for values of the bias current corresponding to different states observed in free-running operation (a fundamental comb in this case, and a CW emission in the case of Fig.~\ref{Figure2}).\\
One notable difference compared to the map shown in Fig.\ref{Figure2}(a) is the possibility to achieve harmonic comb operation and harmonic order tunability for low values of the feedback ratio. For instance, for $\epsilon^2=1~\%$, we report a sequence of OFCs and HFCs with different order in the regime of short cavity, as depicted in Fig.~\ref{Figure5}(b). This implies that if the bias current of the laser is set in correspondence of the emission of a pre-existing free-running OFC, the external feedback allows to manipulate the original comb and obtain HFCs with orders determined by the feedback conditions. In the case where the bias current corresponds to a free-running CW emission, this tunability of the harmonic order was observed for higher values of the feedback ratio ($\epsilon^2>10\%$), which may pose challenges in experimental realization. Thus, we conclude that feedback can be readily employed as a tool for comb manipulation, enabling the generation of different types of HFCs when the bias current corresponds to a free-running OFC.\\
Furthermore, even in the case of $I=1.5I_\mathrm{thr}$, corresponding to comb emission in free-running operation, we have considered how a comb regime is affected by variations in the length of the external cavity on the wavelength scale. In this case, we report an alternation between comb and irregular dynamics in the short cavity regime, and an alternation between comb and mixed states in the long cavity regime, with comb emission periodically obtained with a period of $\lambda$/2. These results are in agreement with the experiments reported in ref.~\cite{Liao22}. Further details are provided in section S.5 of the supplementary material. We notice that some differences occur in the long cavity regime with respect to the case $I=1.08I_\mathrm{thr}$, previously discussed in Sec.~\ref{SecCW} and in Sec. S.2 of the supplementary materials, where it was shown that the comb regimes are not affected by the tuning of $L_\mathrm{ext}$ on the wavelength scale. Therefore, we highlight that according to our numerical results, in the long cavity regime the behaviour of the comb emission under fine-tuning of the external cavity length depends on the value of the bias current, and the results are more stable if the bias current corresponds to single mode emission in free-running operation.
\section{Conclusion}\label{SecConcl}
In conclusion, we conducted a study on the interactions between frequency combs and optical feedback effects in QCLs. Our theoretical analysis predicted new phenomena that arise from the interplay between comb generation and feedback, offering novel solutions with potential applications in various domains. Firstly, we observed that when the bias current corresponds to free-running single mode emission, the introduction of optical feedback can trigger the generation of frequency combs, including both fundamental and harmonic combs. This provides opportunities to extend the comb region in terms of bias current and to generate harmonic frequency combs with different orders by optimizing the external cavity parameters such as losses and length. Additionally, we demonstrated that if a pre-existing free-running comb is emitted by the QCL, the optical feedback can be used to selectively tune its harmonic order by adjusting the external cavity length, particularly for feedback ratios around $1\%$, which are readily achievable in experimental setups. Lastly, under strong feedback conditions (Acket parameter $C>4.6$), we observed the emergence of mixed states with timescales influenced by both the laser and external cavity dynamics. The dynamical scenario in the presence of feedback was also explored for different values of the $\alpha$ factor and carrier lifetime in QCLs, highlighting that the newly discovered phenomena can be observed in both mid-IR and THz QCLs.\\ 
The ability to manipulate and control frequency combs through optical feedback provides unprecedented opportunities for developing innovative technologies with enhanced performance and versatility. The tunability of comb sources enables precise spectral control, facilitating the intentional generation of harmonic states. This capability has significant implications for broadband spectroscopy, sensing, and free-space communication, which are the most compelling applications of the harmonic combs. Moreover, the integration of comb technology with optical feedback offers increased dynamic range, and enhanced spectral coverage. Thus, the novel phenomena reported in our manuscript hold great promise for driving advancements in a wide range of technological domains, such as design of tunable comb sources, hyperspectral imaging, coherent sensing, and multi-channel communication applications.\\
\section*{Supplementary Material}
The supplementary material describes the ESMBEs with optical feedback, the parameters used in the simulations and the procedure followed to classify the dinamical regimes, the behaviour of the frequency combs under fine-tuning of the external cavity length, and an estimation of the critical value of $\alpha$ factor at which irregular regimes arise. It also presents the feedback diagram for a mid-IR QCL. 


\section*{Author Declarations}
The authors have no conflicts to disclose.
\section*{Data Availability}
The data that support the findings of this study are available from the corresponding author upon reasonable request. 
\section*{References}

\bibliography{aipsamp}

\begin{thebibliography}{48}%
\makeatletter
\providecommand \@ifxundefined [1]{%
 \@ifx{#1\undefined}
}%
\providecommand \@ifnum [1]{%
 \ifnum #1\expandafter \@firstoftwo
 \else \expandafter \@secondoftwo
 \fi
}%
\providecommand \@ifx [1]{%
 \ifx #1\expandafter \@firstoftwo
 \else \expandafter \@secondoftwo
 \fi
}%
\providecommand \natexlab [1]{#1}%
\providecommand \enquote  [1]{``#1''}%
\providecommand \bibnamefont  [1]{#1}%
\providecommand \bibfnamefont [1]{#1}%
\providecommand \citenamefont [1]{#1}%
\providecommand \href@noop [0]{\@secondoftwo}%
\providecommand \href [0]{\begingroup \@sanitize@url \@href}%
\providecommand \@href[1]{\@@startlink{#1}\@@href}%
\providecommand \@@href[1]{\endgroup#1\@@endlink}%
\providecommand \@sanitize@url [0]{\catcode `\\12\catcode `\$12\catcode `\&12\catcode `\#12\catcode `\^12\catcode `\_12\catcode `\%12\relax}%
\providecommand \@@startlink[1]{}%
\providecommand \@@endlink[0]{}%
\providecommand \url  [0]{\begingroup\@sanitize@url \@url }%
\providecommand \@url [1]{\endgroup\@href {#1}{\urlprefix }}%
\providecommand \urlprefix  [0]{URL }%
\providecommand \Eprint [0]{\href }%
\providecommand \doibase [0]{http://dx.doi.org/}%
\providecommand \selectlanguage [0]{\@gobble}%
\providecommand \bibinfo  [0]{\@secondoftwo}%
\providecommand \bibfield  [0]{\@secondoftwo}%
\providecommand \translation [1]{[#1]}%
\providecommand \BibitemOpen [0]{}%
\providecommand \bibitemStop [0]{}%
\providecommand \bibitemNoStop [0]{.\EOS\space}%
\providecommand \EOS [0]{\spacefactor3000\relax}%
\providecommand \BibitemShut  [1]{\csname bibitem#1\endcsname}%
\let\auto@bib@innerbib\@empty
\bibitem [{\citenamefont {Faist}\ \emph {et~al.}(1994)\citenamefont {Faist}, \citenamefont {Capasso}, \citenamefont {Sivco}, \citenamefont {Sirtori}, \citenamefont {Hutchinson},\ and\ \citenamefont {Cho}}]{Faist_1994}%
  \BibitemOpen
  \bibfield  {author} {\bibinfo {author} {\bibfnamefont {J.}~\bibnamefont {Faist}}, \bibinfo {author} {\bibfnamefont {F.}~\bibnamefont {Capasso}}, \bibinfo {author} {\bibfnamefont {D.~L.}\ \bibnamefont {Sivco}}, \bibinfo {author} {\bibfnamefont {C.}~\bibnamefont {Sirtori}}, \bibinfo {author} {\bibfnamefont {A.~L.}\ \bibnamefont {Hutchinson}}, \ and\ \bibinfo {author} {\bibfnamefont {A.~Y.}\ \bibnamefont {Cho}},\ }\bibfield  {title} {\enquote {\bibinfo {title} {Quantum cascade laser},}\ }\href {\doibase 10.1126/science.264.5158.553} {\bibfield  {journal} {\bibinfo  {journal} {Science}\ }\textbf {\bibinfo {volume} {264}},\ \bibinfo {pages} {553--556} (\bibinfo {year} {1994})}\BibitemShut {NoStop}%
\bibitem [{\citenamefont {Faist}(2013)}]{Faist_book}%
  \BibitemOpen
  \bibfield  {author} {\bibinfo {author} {\bibfnamefont {J.}~\bibnamefont {Faist}},\ }\href {\doibase 10.1093/acprof:oso/9780198528241.001.0001} {\emph {\bibinfo {title} {Quantum Cascade Lasers}}}\ (\bibinfo  {publisher} {Oxford University Press},\ \bibinfo {year} {2013})\BibitemShut {NoStop}%
\bibitem [{\citenamefont {Vitiello}\ \emph {et~al.}(2015)\citenamefont {Vitiello}, \citenamefont {Scalari}, \citenamefont {Williams},\ and\ \citenamefont {Natale}}]{Vitiellorev}%
  \BibitemOpen
  \bibfield  {author} {\bibinfo {author} {\bibfnamefont {M.~S.}\ \bibnamefont {Vitiello}}, \bibinfo {author} {\bibfnamefont {G.}~\bibnamefont {Scalari}}, \bibinfo {author} {\bibfnamefont {B.}~\bibnamefont {Williams}}, \ and\ \bibinfo {author} {\bibfnamefont {P.~D.}\ \bibnamefont {Natale}},\ }\bibfield  {title} {\enquote {\bibinfo {title} {Quantum cascade lasers: 20 years of challenges},}\ }\href {\doibase 10.1364/OE.23.005167} {\bibfield  {journal} {\bibinfo  {journal} {Opt. Express}\ }\textbf {\bibinfo {volume} {23}},\ \bibinfo {pages} {5167--5182} (\bibinfo {year} {2015})}\BibitemShut {NoStop}%
\bibitem [{\citenamefont {K{\"o}hler}\ \emph {et~al.}(2002)\citenamefont {K{\"o}hler}, \citenamefont {Tredicucci}, \citenamefont {Beltram}, \citenamefont {Beere}, \citenamefont {Linfield}, \citenamefont {Davies}, \citenamefont {Ritchie}, \citenamefont {Iotti},\ and\ \citenamefont {Rossi}}]{Kohler2002}%
  \BibitemOpen
  \bibfield  {author} {\bibinfo {author} {\bibfnamefont {R.}~\bibnamefont {K{\"o}hler}}, \bibinfo {author} {\bibfnamefont {A.}~\bibnamefont {Tredicucci}}, \bibinfo {author} {\bibfnamefont {F.}~\bibnamefont {Beltram}}, \bibinfo {author} {\bibfnamefont {H.~E.}\ \bibnamefont {Beere}}, \bibinfo {author} {\bibfnamefont {E.~H.}\ \bibnamefont {Linfield}}, \bibinfo {author} {\bibfnamefont {A.~G.}\ \bibnamefont {Davies}}, \bibinfo {author} {\bibfnamefont {D.~A.}\ \bibnamefont {Ritchie}}, \bibinfo {author} {\bibfnamefont {R.~C.}\ \bibnamefont {Iotti}}, \ and\ \bibinfo {author} {\bibfnamefont {F.}~\bibnamefont {Rossi}},\ }\bibfield  {title} {\enquote {\bibinfo {title} {Terahertz semiconductor-heterostructure laser},}\ }\href {\doibase https://doi.org/10.1038/417156a} {\bibfield  {journal} {\bibinfo  {journal} {Nature}\ }\textbf {\bibinfo {volume} {417}},\ \bibinfo {pages} {156--159} (\bibinfo {year} {2002})}\BibitemShut {NoStop}%
\bibitem [{\citenamefont {Hugi}\ \emph {et~al.}(2012)\citenamefont {Hugi}, \citenamefont {Villares}, \citenamefont {Blaser}, \citenamefont {Liu},\ and\ \citenamefont {Faist}}]{Hugi2012}%
  \BibitemOpen
  \bibfield  {author} {\bibinfo {author} {\bibfnamefont {A.}~\bibnamefont {Hugi}}, \bibinfo {author} {\bibfnamefont {G.}~\bibnamefont {Villares}}, \bibinfo {author} {\bibfnamefont {S.}~\bibnamefont {Blaser}}, \bibinfo {author} {\bibfnamefont {H.~C.}\ \bibnamefont {Liu}}, \ and\ \bibinfo {author} {\bibfnamefont {J.}~\bibnamefont {Faist}},\ }\bibfield  {title} {\enquote {\bibinfo {title} {Mid-infrared frequency comb based on a quantum cascade laser},}\ }\href {\doibase 10.1038/nature11620} {\bibfield  {journal} {\bibinfo  {journal} {Nature}\ }\textbf {\bibinfo {volume} {492}},\ \bibinfo {pages} {229--233} (\bibinfo {year} {2012})}\BibitemShut {NoStop}%
\bibitem [{\citenamefont {Faist}\ \emph {et~al.}(2016)\citenamefont {Faist}, \citenamefont {Villares}, \citenamefont {Scalari}, \citenamefont {Rösch}, \citenamefont {Bonzon}, \citenamefont {Hugi},\ and\ \citenamefont {Beck}}]{Faist_2016}%
  \BibitemOpen
  \bibfield  {author} {\bibinfo {author} {\bibfnamefont {J.}~\bibnamefont {Faist}}, \bibinfo {author} {\bibfnamefont {G.}~\bibnamefont {Villares}}, \bibinfo {author} {\bibfnamefont {G.}~\bibnamefont {Scalari}}, \bibinfo {author} {\bibfnamefont {M.}~\bibnamefont {Rösch}}, \bibinfo {author} {\bibfnamefont {C.}~\bibnamefont {Bonzon}}, \bibinfo {author} {\bibfnamefont {A.}~\bibnamefont {Hugi}}, \ and\ \bibinfo {author} {\bibfnamefont {M.}~\bibnamefont {Beck}},\ }\bibfield  {title} {\enquote {\bibinfo {title} {Quantum cascade laser frequency combs},}\ }\href {\doibase doi:10.1515/nanoph-2016-0015} {\bibfield  {journal} {\bibinfo  {journal} {Nanophotonics}\ }\textbf {\bibinfo {volume} {5}},\ \bibinfo {pages} {272--291} (\bibinfo {year} {2016})}\BibitemShut {NoStop}%
\bibitem [{\citenamefont {Burghoff}\ \emph {et~al.}(2014)\citenamefont {Burghoff}, \citenamefont {Kao}, \citenamefont {Han}, \citenamefont {Chan}, \citenamefont {Cai}, \citenamefont {Yang}, \citenamefont {Hayton}, \citenamefont {Gao}, \citenamefont {Reno},\ and\ \citenamefont {Hu}}]{Burghoff2014}%
  \BibitemOpen
  \bibfield  {author} {\bibinfo {author} {\bibfnamefont {D.}~\bibnamefont {Burghoff}}, \bibinfo {author} {\bibfnamefont {T.-Y.}\ \bibnamefont {Kao}}, \bibinfo {author} {\bibfnamefont {N.}~\bibnamefont {Han}}, \bibinfo {author} {\bibfnamefont {C.~W.~I.}\ \bibnamefont {Chan}}, \bibinfo {author} {\bibfnamefont {X.}~\bibnamefont {Cai}}, \bibinfo {author} {\bibfnamefont {Y.}~\bibnamefont {Yang}}, \bibinfo {author} {\bibfnamefont {D.~J.}\ \bibnamefont {Hayton}}, \bibinfo {author} {\bibfnamefont {J.-R.}\ \bibnamefont {Gao}}, \bibinfo {author} {\bibfnamefont {J.~L.}\ \bibnamefont {Reno}}, \ and\ \bibinfo {author} {\bibfnamefont {Q.}~\bibnamefont {Hu}},\ }\bibfield  {title} {\enquote {\bibinfo {title} {Terahertz laser frequency combs},}\ }\href {\doibase 10.1038/nphoton.2014.85} {\bibfield  {journal} {\bibinfo  {journal} {Nature Photonics}\ }\textbf {\bibinfo {volume} {8}},\ \bibinfo {pages} {462--467} (\bibinfo {year} {2014})}\BibitemShut {NoStop}%
\bibitem [{\citenamefont {Silvestri}\ \emph {et~al.}(2023)\citenamefont {Silvestri}, \citenamefont {Qi}, \citenamefont {Taimre}, \citenamefont {Bertling},\ and\ \citenamefont {Rakić}}]{SilvestriReview}%
  \BibitemOpen
  \bibfield  {author} {\bibinfo {author} {\bibfnamefont {C.}~\bibnamefont {Silvestri}}, \bibinfo {author} {\bibfnamefont {X.}~\bibnamefont {Qi}}, \bibinfo {author} {\bibfnamefont {T.}~\bibnamefont {Taimre}}, \bibinfo {author} {\bibfnamefont {K.}~\bibnamefont {Bertling}}, \ and\ \bibinfo {author} {\bibfnamefont {A.~D.}\ \bibnamefont {Rakić}},\ }\bibfield  {title} {\enquote {\bibinfo {title} {Frequency combs in quantum cascade lasers: An overview of modeling and experiments},}\ }\href {\doibase https://doi.org/10.1063/5.0134539} {\bibfield  {journal} {\bibinfo  {journal} {APL Photonics}\ }\textbf {\bibinfo {volume} {8}},\ \bibinfo {pages} {020902} (\bibinfo {year} {2023})}\BibitemShut {NoStop}%
\bibitem [{\citenamefont {Piccardo}\ and\ \citenamefont {Capasso}(2022)}]{PiccardoReview}%
  \BibitemOpen
  \bibfield  {author} {\bibinfo {author} {\bibfnamefont {M.}~\bibnamefont {Piccardo}}\ and\ \bibinfo {author} {\bibfnamefont {F.}~\bibnamefont {Capasso}},\ }\bibfield  {title} {\enquote {\bibinfo {title} {Laser frequency combs with fast gain recovery: Physics and applications},}\ }\href {\doibase https://doi.org/10.1002/lpor.202100403} {\bibfield  {journal} {\bibinfo  {journal} {Laser \& Photonics Reviews}\ }\textbf {\bibinfo {volume} {16}},\ \bibinfo {pages} {2100403} (\bibinfo {year} {2022})}\BibitemShut {NoStop}%
\bibitem [{\citenamefont {Opacak}\ and\ \citenamefont {Schwarz}(2019)}]{Opacak2019}%
  \BibitemOpen
  \bibfield  {author} {\bibinfo {author} {\bibfnamefont {N.}~\bibnamefont {Opacak}}\ and\ \bibinfo {author} {\bibfnamefont {B.}~\bibnamefont {Schwarz}},\ }\bibfield  {title} {\enquote {\bibinfo {title} {Theory of frequency-modulated combs in lasers with spatial hole burning, dispersion, and kerr nonlinearity},}\ }\href {\doibase 10.1103/PhysRevLett.123.243902} {\bibfield  {journal} {\bibinfo  {journal} {Phys. Rev. Lett.}\ }\textbf {\bibinfo {volume} {123}},\ \bibinfo {pages} {243902} (\bibinfo {year} {2019})}\BibitemShut {NoStop}%
\bibitem [{\citenamefont {Singleton}\ \emph {et~al.}(2018)\citenamefont {Singleton}, \citenamefont {Jouy}, \citenamefont {Beck},\ and\ \citenamefont {Faist}}]{Singleton18}%
  \BibitemOpen
  \bibfield  {author} {\bibinfo {author} {\bibfnamefont {M.}~\bibnamefont {Singleton}}, \bibinfo {author} {\bibfnamefont {P.}~\bibnamefont {Jouy}}, \bibinfo {author} {\bibfnamefont {M.}~\bibnamefont {Beck}}, \ and\ \bibinfo {author} {\bibfnamefont {J.}~\bibnamefont {Faist}},\ }\bibfield  {title} {\enquote {\bibinfo {title} {Evidence of linear chirp in mid-infrared quantum cascade lasers},}\ }\href {\doibase 10.1364/OPTICA.5.000948} {\bibfield  {journal} {\bibinfo  {journal} {Optica}\ }\textbf {\bibinfo {volume} {5}},\ \bibinfo {pages} {948--953} (\bibinfo {year} {2018})}\BibitemShut {NoStop}%
\bibitem [{\citenamefont {Columbo}\ \emph {et~al.}(2021)\citenamefont {Columbo}, \citenamefont {Piccardo}, \citenamefont {Prati}, \citenamefont {Lugiato}, \citenamefont {Brambilla}, \citenamefont {Gatti}, \citenamefont {Silvestri}, \citenamefont {Gioannini}, \citenamefont {Opacak}, \citenamefont {Schwarz},\ and\ \citenamefont {Capasso}}]{Unifying}%
  \BibitemOpen
  \bibfield  {author} {\bibinfo {author} {\bibfnamefont {L.}~\bibnamefont {Columbo}}, \bibinfo {author} {\bibfnamefont {M.}~\bibnamefont {Piccardo}}, \bibinfo {author} {\bibfnamefont {F.}~\bibnamefont {Prati}}, \bibinfo {author} {\bibfnamefont {L.~A.}\ \bibnamefont {Lugiato}}, \bibinfo {author} {\bibfnamefont {M.}~\bibnamefont {Brambilla}}, \bibinfo {author} {\bibfnamefont {A.}~\bibnamefont {Gatti}}, \bibinfo {author} {\bibfnamefont {C.}~\bibnamefont {Silvestri}}, \bibinfo {author} {\bibfnamefont {M.}~\bibnamefont {Gioannini}}, \bibinfo {author} {\bibfnamefont {N.}~\bibnamefont {Opacak}}, \bibinfo {author} {\bibfnamefont {B.}~\bibnamefont {Schwarz}}, \ and\ \bibinfo {author} {\bibfnamefont {F.}~\bibnamefont {Capasso}},\ }\bibfield  {title} {\enquote {\bibinfo {title} {Unifying frequency combs in active and passive cavities: Temporal solitons in externally driven ring lasers},}\ }\href {\doibase 10.1103/PhysRevLett.126.173903} {\bibfield  {journal} {\bibinfo  {journal} {Phys. Rev. Lett.}\ }\textbf {\bibinfo
  {volume} {126}},\ \bibinfo {pages} {173903} (\bibinfo {year} {2021})}\BibitemShut {NoStop}%
\bibitem [{\citenamefont {Prati}\ \emph {et~al.}(2021{\natexlab{a}})\citenamefont {Prati}, \citenamefont {Lugiato}, \citenamefont {Gatti}, \citenamefont {Columbo}, \citenamefont {Silvestri}, \citenamefont {Gioannini}, \citenamefont {Brambilla}, \citenamefont {Piccardo},\ and\ \citenamefont {Capasso}}]{Pratichaos}%
  \BibitemOpen
  \bibfield  {author} {\bibinfo {author} {\bibfnamefont {F.}~\bibnamefont {Prati}}, \bibinfo {author} {\bibfnamefont {L.}~\bibnamefont {Lugiato}}, \bibinfo {author} {\bibfnamefont {A.}~\bibnamefont {Gatti}}, \bibinfo {author} {\bibfnamefont {L.}~\bibnamefont {Columbo}}, \bibinfo {author} {\bibfnamefont {C.}~\bibnamefont {Silvestri}}, \bibinfo {author} {\bibfnamefont {M.}~\bibnamefont {Gioannini}}, \bibinfo {author} {\bibfnamefont {M.}~\bibnamefont {Brambilla}}, \bibinfo {author} {\bibfnamefont {M.}~\bibnamefont {Piccardo}}, \ and\ \bibinfo {author} {\bibfnamefont {F.}~\bibnamefont {Capasso}},\ }\bibfield  {title} {\enquote {\bibinfo {title} {Global and localised temporal structures in driven ring quantum cascade lasers},}\ }\href {\doibase https://doi.org/10.1016/j.chaos.2021.111537} {\bibfield  {journal} {\bibinfo  {journal} {Chaos, Solitons $\&$ Fractals}\ }\textbf {\bibinfo {volume} {153}},\ \bibinfo {pages} {111537} (\bibinfo {year} {2021}{\natexlab{a}})}\BibitemShut {NoStop}%
\bibitem [{\citenamefont {Prati}\ \emph {et~al.}(2021{\natexlab{b}})\citenamefont {Prati}, \citenamefont {Brambilla}, \citenamefont {Piccardo}, \citenamefont {Columbo}, \citenamefont {Silvestri}, \citenamefont {Gioannini}, \citenamefont {Gatti}, \citenamefont {Lugiato},\ and\ \citenamefont {Capasso}}]{Pratinano}%
  \BibitemOpen
  \bibfield  {author} {\bibinfo {author} {\bibfnamefont {F.}~\bibnamefont {Prati}}, \bibinfo {author} {\bibfnamefont {M.}~\bibnamefont {Brambilla}}, \bibinfo {author} {\bibfnamefont {M.}~\bibnamefont {Piccardo}}, \bibinfo {author} {\bibfnamefont {L.~L.}\ \bibnamefont {Columbo}}, \bibinfo {author} {\bibfnamefont {C.}~\bibnamefont {Silvestri}}, \bibinfo {author} {\bibfnamefont {M.}~\bibnamefont {Gioannini}}, \bibinfo {author} {\bibfnamefont {A.}~\bibnamefont {Gatti}}, \bibinfo {author} {\bibfnamefont {L.~A.}\ \bibnamefont {Lugiato}}, \ and\ \bibinfo {author} {\bibfnamefont {F.}~\bibnamefont {Capasso}},\ }\bibfield  {title} {\enquote {\bibinfo {title} {Soliton dynamics of ring quantum cascade lasers with injected signal},}\ }\href {\doibase doi:10.1515/nanoph-2020-0409} {\bibfield  {journal} {\bibinfo  {journal} {Nanophotonics}\ }\textbf {\bibinfo {volume} {10}},\ \bibinfo {pages} {195--207} (\bibinfo {year} {2021}{\natexlab{b}})}\BibitemShut {NoStop}%
\bibitem [{\citenamefont {Piccardo}\ \emph {et~al.}(2020)\citenamefont {Piccardo}, \citenamefont {Schwarz}, \citenamefont {Kazakov}, \citenamefont {Beiser}, \citenamefont {Opa{\v{c}}ak}, \citenamefont {Wang}, \citenamefont {Jha}, \citenamefont {Hillbrand}, \citenamefont {Tamagnone}, \citenamefont {Chen}, \citenamefont {Zhu}, \citenamefont {Columbo}, \citenamefont {Belyanin},\ and\ \citenamefont {Capasso}}]{NaturePiccardo}%
  \BibitemOpen
  \bibfield  {author} {\bibinfo {author} {\bibfnamefont {M.}~\bibnamefont {Piccardo}}, \bibinfo {author} {\bibfnamefont {B.}~\bibnamefont {Schwarz}}, \bibinfo {author} {\bibfnamefont {D.}~\bibnamefont {Kazakov}}, \bibinfo {author} {\bibfnamefont {M.}~\bibnamefont {Beiser}}, \bibinfo {author} {\bibfnamefont {N.}~\bibnamefont {Opa{\v{c}}ak}}, \bibinfo {author} {\bibfnamefont {Y.}~\bibnamefont {Wang}}, \bibinfo {author} {\bibfnamefont {S.}~\bibnamefont {Jha}}, \bibinfo {author} {\bibfnamefont {J.}~\bibnamefont {Hillbrand}}, \bibinfo {author} {\bibfnamefont {M.}~\bibnamefont {Tamagnone}}, \bibinfo {author} {\bibfnamefont {W.~T.}\ \bibnamefont {Chen}}, \bibinfo {author} {\bibfnamefont {A.~Y.}\ \bibnamefont {Zhu}}, \bibinfo {author} {\bibfnamefont {L.~L.}\ \bibnamefont {Columbo}}, \bibinfo {author} {\bibfnamefont {A.}~\bibnamefont {Belyanin}}, \ and\ \bibinfo {author} {\bibfnamefont {F.}~\bibnamefont {Capasso}},\ }\bibfield  {title} {\enquote {\bibinfo {title} {Frequency combs induced by phase turbulence},}\ }\href
  {\doibase 10.1038/s41586-020-2386-6} {\bibfield  {journal} {\bibinfo  {journal} {Nature}\ }\textbf {\bibinfo {volume} {582}},\ \bibinfo {pages} {360--364} (\bibinfo {year} {2020})}\BibitemShut {NoStop}%
\bibitem [{\citenamefont {Meng}\ \emph {et~al.}(2022)\citenamefont {Meng}, \citenamefont {Singleton}, \citenamefont {Hillbrand}, \citenamefont {Francki{\'e}}, \citenamefont {Beck},\ and\ \citenamefont {Faist}}]{Bomeng2}%
  \BibitemOpen
  \bibfield  {author} {\bibinfo {author} {\bibfnamefont {B.}~\bibnamefont {Meng}}, \bibinfo {author} {\bibfnamefont {M.}~\bibnamefont {Singleton}}, \bibinfo {author} {\bibfnamefont {J.}~\bibnamefont {Hillbrand}}, \bibinfo {author} {\bibfnamefont {M.}~\bibnamefont {Francki{\'e}}}, \bibinfo {author} {\bibfnamefont {M.}~\bibnamefont {Beck}}, \ and\ \bibinfo {author} {\bibfnamefont {J.}~\bibnamefont {Faist}},\ }\bibfield  {title} {\enquote {\bibinfo {title} {Dissipative {K}err solitons in semiconductor ring lasers},}\ }\href {\doibase https://doi.org/10.1038/s41566-021-00927-3} {\bibfield  {journal} {\bibinfo  {journal} {Nature Photonics}\ }\textbf {\bibinfo {volume} {16}},\ \bibinfo {pages} {142--147} (\bibinfo {year} {2022})}\BibitemShut {NoStop}%
\bibitem [{\citenamefont {Schneider}\ \emph {et~al.}(2021)\citenamefont {Schneider}, \citenamefont {Kapsalidis}, \citenamefont {Bertrand}, \citenamefont {Singleton}, \citenamefont {Hillbrand}, \citenamefont {Beck},\ and\ \citenamefont {Faist}}]{schneider}%
  \BibitemOpen
  \bibfield  {author} {\bibinfo {author} {\bibfnamefont {B.}~\bibnamefont {Schneider}}, \bibinfo {author} {\bibfnamefont {F.}~\bibnamefont {Kapsalidis}}, \bibinfo {author} {\bibfnamefont {M.}~\bibnamefont {Bertrand}}, \bibinfo {author} {\bibfnamefont {M.}~\bibnamefont {Singleton}}, \bibinfo {author} {\bibfnamefont {J.}~\bibnamefont {Hillbrand}}, \bibinfo {author} {\bibfnamefont {M.}~\bibnamefont {Beck}}, \ and\ \bibinfo {author} {\bibfnamefont {J.}~\bibnamefont {Faist}},\ }\bibfield  {title} {\enquote {\bibinfo {title} {Controlling quantum cascade laser optical frequency combs through microwave injection},}\ }\href {\doibase https://doi.org/10.1002/lpor.202100242} {\bibfield  {journal} {\bibinfo  {journal} {Laser \& Photonics Reviews}\ }\textbf {\bibinfo {volume} {15}},\ \bibinfo {pages} {2100242} (\bibinfo {year} {2021})}\BibitemShut {NoStop}%
\bibitem [{\citenamefont {Consolino}\ \emph {et~al.}(2021)\citenamefont {Consolino}, \citenamefont {Campa}, \citenamefont {De~Regis}, \citenamefont {Cappelli}, \citenamefont {Scalari}, \citenamefont {Faist}, \citenamefont {Beck}, \citenamefont {Rösch}, \citenamefont {Bartalini},\ and\ \citenamefont {De~Natale}}]{consolino1}%
  \BibitemOpen
  \bibfield  {author} {\bibinfo {author} {\bibfnamefont {L.}~\bibnamefont {Consolino}}, \bibinfo {author} {\bibfnamefont {A.}~\bibnamefont {Campa}}, \bibinfo {author} {\bibfnamefont {M.}~\bibnamefont {De~Regis}}, \bibinfo {author} {\bibfnamefont {F.}~\bibnamefont {Cappelli}}, \bibinfo {author} {\bibfnamefont {G.}~\bibnamefont {Scalari}}, \bibinfo {author} {\bibfnamefont {J.}~\bibnamefont {Faist}}, \bibinfo {author} {\bibfnamefont {M.}~\bibnamefont {Beck}}, \bibinfo {author} {\bibfnamefont {M.}~\bibnamefont {Rösch}}, \bibinfo {author} {\bibfnamefont {S.}~\bibnamefont {Bartalini}}, \ and\ \bibinfo {author} {\bibfnamefont {P.}~\bibnamefont {De~Natale}},\ }\bibfield  {title} {\enquote {\bibinfo {title} {Controlling and phase-locking a thz quantum cascade laser frequency comb by small optical frequency tuning},}\ }\href {\doibase https://doi.org/10.1002/lpor.202000417} {\bibfield  {journal} {\bibinfo  {journal} {Laser \& Photonics Reviews}\ }\textbf {\bibinfo {volume} {15}},\ \bibinfo {pages} {2000417} (\bibinfo
  {year} {2021})}\BibitemShut {NoStop}%
\bibitem [{\citenamefont {Piccardo}\ \emph {et~al.}(2018{\natexlab{a}})\citenamefont {Piccardo}, \citenamefont {Chevalier}, \citenamefont {Anand}, \citenamefont {Wang}, \citenamefont {Kazakov}, \citenamefont {Mejia}, \citenamefont {Xie}, \citenamefont {Lascola}, \citenamefont {Belyanin},\ and\ \citenamefont {Capasso}}]{PiccardoOptical}%
  \BibitemOpen
  \bibfield  {author} {\bibinfo {author} {\bibfnamefont {M.}~\bibnamefont {Piccardo}}, \bibinfo {author} {\bibfnamefont {P.}~\bibnamefont {Chevalier}}, \bibinfo {author} {\bibfnamefont {S.}~\bibnamefont {Anand}}, \bibinfo {author} {\bibfnamefont {Y.}~\bibnamefont {Wang}}, \bibinfo {author} {\bibfnamefont {D.}~\bibnamefont {Kazakov}}, \bibinfo {author} {\bibfnamefont {E.~A.}\ \bibnamefont {Mejia}}, \bibinfo {author} {\bibfnamefont {F.}~\bibnamefont {Xie}}, \bibinfo {author} {\bibfnamefont {K.}~\bibnamefont {Lascola}}, \bibinfo {author} {\bibfnamefont {A.}~\bibnamefont {Belyanin}}, \ and\ \bibinfo {author} {\bibfnamefont {F.}~\bibnamefont {Capasso}},\ }\bibfield  {title} {\enquote {\bibinfo {title} {Widely tunable harmonic frequency comb in a quantum cascade laser},}\ }\href {\doibase https://doi.org/10.1063/1.5039611} {\bibfield  {journal} {\bibinfo  {journal} {Applied Physics Letters}\ }\textbf {\bibinfo {volume} {113}},\ \bibinfo {pages} {031104} (\bibinfo {year} {2018}{\natexlab{a}})}\BibitemShut {NoStop}%
\bibitem [{\citenamefont {Rakić}\ \emph {et~al.}(2019)\citenamefont {Rakić}, \citenamefont {Taimre}, \citenamefont {Bertling}, \citenamefont {Lim}, \citenamefont {Dean}, \citenamefont {Valavanis},\ and\ \citenamefont {Indjin}}]{Rakicreview}%
  \BibitemOpen
  \bibfield  {author} {\bibinfo {author} {\bibfnamefont {A.~D.}\ \bibnamefont {Rakić}}, \bibinfo {author} {\bibfnamefont {T.}~\bibnamefont {Taimre}}, \bibinfo {author} {\bibfnamefont {K.}~\bibnamefont {Bertling}}, \bibinfo {author} {\bibfnamefont {Y.~L.}\ \bibnamefont {Lim}}, \bibinfo {author} {\bibfnamefont {P.}~\bibnamefont {Dean}}, \bibinfo {author} {\bibfnamefont {A.}~\bibnamefont {Valavanis}}, \ and\ \bibinfo {author} {\bibfnamefont {D.}~\bibnamefont {Indjin}},\ }\bibfield  {title} {\enquote {\bibinfo {title} {Sensing and imaging using laser feedback interferometry with quantum cascade lasers},}\ }\href {\doibase https://doi.org/10.1063/1.5094674} {\bibfield  {journal} {\bibinfo  {journal} {Applied Physics Reviews}\ }\textbf {\bibinfo {volume} {6}},\ \bibinfo {pages} {021320} (\bibinfo {year} {2019})}\BibitemShut {NoStop}%
\bibitem [{\citenamefont {Mezzapesa}\ \emph {et~al.}(2013)\citenamefont {Mezzapesa}, \citenamefont {Columbo}, \citenamefont {Brambilla}, \citenamefont {Dabbicco}, \citenamefont {Borri}, \citenamefont {Vitiello}, \citenamefont {Beere}, \citenamefont {Ritchie},\ and\ \citenamefont {Scamarcio}}]{mezzapesa2013}%
  \BibitemOpen
  \bibfield  {author} {\bibinfo {author} {\bibfnamefont {F.~P.}\ \bibnamefont {Mezzapesa}}, \bibinfo {author} {\bibfnamefont {L.~L.}\ \bibnamefont {Columbo}}, \bibinfo {author} {\bibfnamefont {M.}~\bibnamefont {Brambilla}}, \bibinfo {author} {\bibfnamefont {M.}~\bibnamefont {Dabbicco}}, \bibinfo {author} {\bibfnamefont {S.}~\bibnamefont {Borri}}, \bibinfo {author} {\bibfnamefont {M.~S.}\ \bibnamefont {Vitiello}}, \bibinfo {author} {\bibfnamefont {H.~E.}\ \bibnamefont {Beere}}, \bibinfo {author} {\bibfnamefont {D.~A.}\ \bibnamefont {Ritchie}}, \ and\ \bibinfo {author} {\bibfnamefont {G.}~\bibnamefont {Scamarcio}},\ }\bibfield  {title} {\enquote {\bibinfo {title} {Intrinsic stability of quantum cascade lasers against optical feedback},}\ }\href {\doibase 10.1364/OE.21.013748} {\bibfield  {journal} {\bibinfo  {journal} {Opt. Express}\ }\textbf {\bibinfo {volume} {21}},\ \bibinfo {pages} {13748--13757} (\bibinfo {year} {2013})}\BibitemShut {NoStop}%
\bibitem [{\citenamefont {Taimre}\ \emph {et~al.}(2015)\citenamefont {Taimre}, \citenamefont {Nikoli\'{c}}, \citenamefont {Bertling}, \citenamefont {Lim}, \citenamefont {Bosch},\ and\ \citenamefont {Raki\'{c}}}]{Taimre15}%
  \BibitemOpen
  \bibfield  {author} {\bibinfo {author} {\bibfnamefont {T.}~\bibnamefont {Taimre}}, \bibinfo {author} {\bibfnamefont {M.}~\bibnamefont {Nikoli\'{c}}}, \bibinfo {author} {\bibfnamefont {K.}~\bibnamefont {Bertling}}, \bibinfo {author} {\bibfnamefont {Y.~L.}\ \bibnamefont {Lim}}, \bibinfo {author} {\bibfnamefont {T.}~\bibnamefont {Bosch}}, \ and\ \bibinfo {author} {\bibfnamefont {A.~D.}\ \bibnamefont {Raki\'{c}}},\ }\bibfield  {title} {\enquote {\bibinfo {title} {Laser feedback interferometry: a tutorial on the self-mixing effect for coherent sensing},}\ }\href {\doibase 10.1364/AOP.7.000570} {\bibfield  {journal} {\bibinfo  {journal} {Adv. Opt. Photon.}\ }\textbf {\bibinfo {volume} {7}},\ \bibinfo {pages} {570--631} (\bibinfo {year} {2015})}\BibitemShut {NoStop}%
\bibitem [{\citenamefont {Jumpertz}\ \emph {et~al.}(2014)\citenamefont {Jumpertz}, \citenamefont {Carras}, \citenamefont {Schires},\ and\ \citenamefont {Grillot}}]{Grillot14}%
  \BibitemOpen
  \bibfield  {author} {\bibinfo {author} {\bibfnamefont {L.}~\bibnamefont {Jumpertz}}, \bibinfo {author} {\bibfnamefont {M.}~\bibnamefont {Carras}}, \bibinfo {author} {\bibfnamefont {K.}~\bibnamefont {Schires}}, \ and\ \bibinfo {author} {\bibfnamefont {F.}~\bibnamefont {Grillot}},\ }\bibfield  {title} {\enquote {\bibinfo {title} {Regimes of external optical feedback in 5.6 ${\mu}m$ distributed feedback mid-infrared quantum cascade lasers},}\ }\href {\doibase https://doi.org/10.1063/1.4897274} {\bibfield  {journal} {\bibinfo  {journal} {Applied Physics Letters}\ }\textbf {\bibinfo {volume} {105}},\ \bibinfo {pages} {131112} (\bibinfo {year} {2014})}\BibitemShut {NoStop}%
\bibitem [{\citenamefont {Jumpertz}\ \emph {et~al.}(2016)\citenamefont {Jumpertz}, \citenamefont {Michel}, \citenamefont {Pawlus}, \citenamefont {Elsasser}, \citenamefont {Schires}, \citenamefont {Carras},\ and\ \citenamefont {Grillot}}]{Grillot16}%
  \BibitemOpen
  \bibfield  {author} {\bibinfo {author} {\bibfnamefont {L.}~\bibnamefont {Jumpertz}}, \bibinfo {author} {\bibfnamefont {F.}~\bibnamefont {Michel}}, \bibinfo {author} {\bibfnamefont {R.}~\bibnamefont {Pawlus}}, \bibinfo {author} {\bibfnamefont {W.}~\bibnamefont {Elsasser}}, \bibinfo {author} {\bibfnamefont {K.}~\bibnamefont {Schires}}, \bibinfo {author} {\bibfnamefont {M.}~\bibnamefont {Carras}}, \ and\ \bibinfo {author} {\bibfnamefont {F.}~\bibnamefont {Grillot}},\ }\bibfield  {title} {\enquote {\bibinfo {title} {Measurements of the linewidth enhancement factor of mid-infrared quantum cascade lasers by different optical feedback techniques},}\ }\href {\doibase https://doi.org/10.1063/1.4940767} {\bibfield  {journal} {\bibinfo  {journal} {AIP Advances}\ }\textbf {\bibinfo {volume} {6}},\ \bibinfo {pages} {015212} (\bibinfo {year} {2016})}\BibitemShut {NoStop}%
\bibitem [{\citenamefont {Pogna}\ \emph {et~al.}(2021)\citenamefont {Pogna}, \citenamefont {Silvestri}, \citenamefont {Columbo}, \citenamefont {Brambilla}, \citenamefont {Scamarcio},\ and\ \citenamefont {Vitiello}}]{Pogna2021}%
  \BibitemOpen
  \bibfield  {author} {\bibinfo {author} {\bibfnamefont {E.~A.~A.}\ \bibnamefont {Pogna}}, \bibinfo {author} {\bibfnamefont {C.}~\bibnamefont {Silvestri}}, \bibinfo {author} {\bibfnamefont {L.~L.}\ \bibnamefont {Columbo}}, \bibinfo {author} {\bibfnamefont {M.}~\bibnamefont {Brambilla}}, \bibinfo {author} {\bibfnamefont {G.}~\bibnamefont {Scamarcio}}, \ and\ \bibinfo {author} {\bibfnamefont {M.~S.}\ \bibnamefont {Vitiello}},\ }\bibfield  {title} {\enquote {\bibinfo {title} {Terahertz near-field nanoscopy based on detectorless laser feedback interferometry under different feedback regimes},}\ }\href {\doibase https://doi.org/10.1063/5.0048099} {\bibfield  {journal} {\bibinfo  {journal} {APL Photonics}\ }\textbf {\bibinfo {volume} {6}},\ \bibinfo {pages} {061302} (\bibinfo {year} {2021})}\BibitemShut {NoStop}%
\bibitem [{\citenamefont {Giordano}\ \emph {et~al.}(2018)\citenamefont {Giordano}, \citenamefont {Mastel}, \citenamefont {Liewald}, \citenamefont {Columbo}, \citenamefont {Brambilla}, \citenamefont {Viti}, \citenamefont {Politano}, \citenamefont {Zhang}, \citenamefont {Li}, \citenamefont {Davies}, \citenamefont {Linfield}, \citenamefont {Hillenbrand}, \citenamefont {Keilmann}, \citenamefont {Scamarcio},\ and\ \citenamefont {Vitiello}}]{Giordano18}%
  \BibitemOpen
  \bibfield  {author} {\bibinfo {author} {\bibfnamefont {M.~C.}\ \bibnamefont {Giordano}}, \bibinfo {author} {\bibfnamefont {S.}~\bibnamefont {Mastel}}, \bibinfo {author} {\bibfnamefont {C.}~\bibnamefont {Liewald}}, \bibinfo {author} {\bibfnamefont {L.~L.}\ \bibnamefont {Columbo}}, \bibinfo {author} {\bibfnamefont {M.}~\bibnamefont {Brambilla}}, \bibinfo {author} {\bibfnamefont {L.}~\bibnamefont {Viti}}, \bibinfo {author} {\bibfnamefont {A.}~\bibnamefont {Politano}}, \bibinfo {author} {\bibfnamefont {K.}~\bibnamefont {Zhang}}, \bibinfo {author} {\bibfnamefont {L.}~\bibnamefont {Li}}, \bibinfo {author} {\bibfnamefont {A.~G.}\ \bibnamefont {Davies}}, \bibinfo {author} {\bibfnamefont {E.~H.}\ \bibnamefont {Linfield}}, \bibinfo {author} {\bibfnamefont {R.}~\bibnamefont {Hillenbrand}}, \bibinfo {author} {\bibfnamefont {F.}~\bibnamefont {Keilmann}}, \bibinfo {author} {\bibfnamefont {G.}~\bibnamefont {Scamarcio}}, \ and\ \bibinfo {author} {\bibfnamefont {M.~S.}\ \bibnamefont {Vitiello}},\ }\bibfield  {title}
  {\enquote {\bibinfo {title} {Phase-resolved terahertz self-detection near-field microscopy},}\ }\href {\doibase 10.1364/OE.26.018423} {\bibfield  {journal} {\bibinfo  {journal} {Opt. Express}\ }\textbf {\bibinfo {volume} {26}},\ \bibinfo {pages} {18423--18435} (\bibinfo {year} {2018})}\BibitemShut {NoStop}%
\bibitem [{\citenamefont {Lim}\ \emph {et~al.}(2019)\citenamefont {Lim}, \citenamefont {Bertling}, \citenamefont {Taimre}, \citenamefont {Gillespie}, \citenamefont {Glenn}, \citenamefont {Robinson}, \citenamefont {Indjin}, \citenamefont {Han}, \citenamefont {Li}, \citenamefont {Linfield}, \citenamefont {Davies}, \citenamefont {Dean},\ and\ \citenamefont {Raki\'{c}}}]{Lim19}%
  \BibitemOpen
  \bibfield  {author} {\bibinfo {author} {\bibfnamefont {Y.~L.}\ \bibnamefont {Lim}}, \bibinfo {author} {\bibfnamefont {K.}~\bibnamefont {Bertling}}, \bibinfo {author} {\bibfnamefont {T.}~\bibnamefont {Taimre}}, \bibinfo {author} {\bibfnamefont {T.}~\bibnamefont {Gillespie}}, \bibinfo {author} {\bibfnamefont {C.}~\bibnamefont {Glenn}}, \bibinfo {author} {\bibfnamefont {A.}~\bibnamefont {Robinson}}, \bibinfo {author} {\bibfnamefont {D.}~\bibnamefont {Indjin}}, \bibinfo {author} {\bibfnamefont {Y.}~\bibnamefont {Han}}, \bibinfo {author} {\bibfnamefont {L.}~\bibnamefont {Li}}, \bibinfo {author} {\bibfnamefont {E.~H.}\ \bibnamefont {Linfield}}, \bibinfo {author} {\bibfnamefont {A.~G.}\ \bibnamefont {Davies}}, \bibinfo {author} {\bibfnamefont {P.}~\bibnamefont {Dean}}, \ and\ \bibinfo {author} {\bibfnamefont {A.~D.}\ \bibnamefont {Raki\'{c}}},\ }\bibfield  {title} {\enquote {\bibinfo {title} {Coherent imaging using laser feedback interferometry with pulsed-mode terahertz quantum cascade lasers},}\ }\href {\doibase
  10.1364/OE.27.010221} {\bibfield  {journal} {\bibinfo  {journal} {Opt. Express}\ }\textbf {\bibinfo {volume} {27}},\ \bibinfo {pages} {10221--10233} (\bibinfo {year} {2019})}\BibitemShut {NoStop}%
\bibitem [{\citenamefont {Silvestri}, \citenamefont {Columbo},\ and\ \citenamefont {Brambilla}(2023)}]{SilvestriSNOM}%
  \BibitemOpen
  \bibfield  {author} {\bibinfo {author} {\bibfnamefont {C.}~\bibnamefont {Silvestri}}, \bibinfo {author} {\bibfnamefont {L.~L.}\ \bibnamefont {Columbo}}, \ and\ \bibinfo {author} {\bibfnamefont {M.}~\bibnamefont {Brambilla}},\ }\bibfield  {title} {\enquote {\bibinfo {title} {Retrieval of the dielectric properties of a resonant material in the terahertz region via self-detection near field optical microscopy},}\ }\href {\doibase 10.1109/JSTQE.2023.3287041} {\bibfield  {journal} {\bibinfo  {journal} {IEEE Journal of Selected Topics in Quantum Electronics}\ ,\ \bibinfo {pages} {1--11}} (\bibinfo {year} {2023})}\BibitemShut {NoStop}%
\bibitem [{\citenamefont {Pistore}\ \emph {et~al.}(2022)\citenamefont {Pistore}, \citenamefont {Pogna}, \citenamefont {Viti}, \citenamefont {Li}, \citenamefont {Davies}, \citenamefont {Linfield},\ and\ \citenamefont {Vitiello}}]{VitielloSNOM2022}%
  \BibitemOpen
  \bibfield  {author} {\bibinfo {author} {\bibfnamefont {V.}~\bibnamefont {Pistore}}, \bibinfo {author} {\bibfnamefont {E.~A.~A.}\ \bibnamefont {Pogna}}, \bibinfo {author} {\bibfnamefont {L.}~\bibnamefont {Viti}}, \bibinfo {author} {\bibfnamefont {L.}~\bibnamefont {Li}}, \bibinfo {author} {\bibfnamefont {A.~G.}\ \bibnamefont {Davies}}, \bibinfo {author} {\bibfnamefont {E.~H.}\ \bibnamefont {Linfield}}, \ and\ \bibinfo {author} {\bibfnamefont {M.~S.}\ \bibnamefont {Vitiello}},\ }\bibfield  {title} {\enquote {\bibinfo {title} {Self-induced phase locking of terahertz frequency combs in a phase-sensitive hyperspectral near-field nanoscope},}\ }\href {\doibase https://doi.org/10.1002/advs.202200410} {\bibfield  {journal} {\bibinfo  {journal} {Advanced Science}\ }\textbf {\bibinfo {volume} {9}},\ \bibinfo {pages} {2200410} (\bibinfo {year} {2022})}\BibitemShut {NoStop}%
\bibitem [{\citenamefont {Teng}, \citenamefont {Westberg},\ and\ \citenamefont {Wysocki}(2019)}]{Teng19}%
  \BibitemOpen
  \bibfield  {author} {\bibinfo {author} {\bibfnamefont {C.~C.}\ \bibnamefont {Teng}}, \bibinfo {author} {\bibfnamefont {J.}~\bibnamefont {Westberg}}, \ and\ \bibinfo {author} {\bibfnamefont {G.}~\bibnamefont {Wysocki}},\ }\bibfield  {title} {\enquote {\bibinfo {title} {Optical-feedback-stabilized quantum cascade laser frequency combs},}\ }in\ \href {\doibase 10.1364/CLEO_SI.2019.STu4N.3} {\emph {\bibinfo {booktitle} {Conference on Lasers and Electro-Optics}}}\ (\bibinfo  {publisher} {Optica Publishing Group},\ \bibinfo {year} {2019})\ p.\ \bibinfo {pages} {STu4N.3}\BibitemShut {NoStop}%
\bibitem [{\citenamefont {Teng}, \citenamefont {Westberg},\ and\ \citenamefont {Wysocki}(2023)}]{Teng23}%
  \BibitemOpen
  \bibfield  {author} {\bibinfo {author} {\bibfnamefont {C.~C.}\ \bibnamefont {Teng}}, \bibinfo {author} {\bibfnamefont {J.}~\bibnamefont {Westberg}}, \ and\ \bibinfo {author} {\bibfnamefont {G.}~\bibnamefont {Wysocki}},\ }\bibfield  {title} {\enquote {\bibinfo {title} {Gapless tuning of quantum cascade laser frequency combs with external cavity optical feedback},}\ }\href {\doibase 10.1364/OL.478950} {\bibfield  {journal} {\bibinfo  {journal} {Opt. Lett.}\ }\textbf {\bibinfo {volume} {48}},\ \bibinfo {pages} {363--366} (\bibinfo {year} {2023})}\BibitemShut {NoStop}%
\bibitem [{\citenamefont {Liao}\ \emph {et~al.}(2022)\citenamefont {Liao}, \citenamefont {Wang}, \citenamefont {Zhou}, \citenamefont {Guan}, \citenamefont {Li}, \citenamefont {Ma}, \citenamefont {Wang}, \citenamefont {Cao}, \citenamefont {Wang},\ and\ \citenamefont {Li}}]{Liao22}%
  \BibitemOpen
  \bibfield  {author} {\bibinfo {author} {\bibfnamefont {X.}~\bibnamefont {Liao}}, \bibinfo {author} {\bibfnamefont {X.}~\bibnamefont {Wang}}, \bibinfo {author} {\bibfnamefont {K.}~\bibnamefont {Zhou}}, \bibinfo {author} {\bibfnamefont {W.}~\bibnamefont {Guan}}, \bibinfo {author} {\bibfnamefont {Z.}~\bibnamefont {Li}}, \bibinfo {author} {\bibfnamefont {X.}~\bibnamefont {Ma}}, \bibinfo {author} {\bibfnamefont {C.}~\bibnamefont {Wang}}, \bibinfo {author} {\bibfnamefont {J.~C.}\ \bibnamefont {Cao}}, \bibinfo {author} {\bibfnamefont {C.}~\bibnamefont {Wang}}, \ and\ \bibinfo {author} {\bibfnamefont {H.}~\bibnamefont {Li}},\ }\bibfield  {title} {\enquote {\bibinfo {title} {Terahertz quantum cascade laser frequency combs with optical feedback},}\ }\href {\doibase 10.1364/OE.467992} {\bibfield  {journal} {\bibinfo  {journal} {Opt. Express}\ }\textbf {\bibinfo {volume} {30}},\ \bibinfo {pages} {35937--35950} (\bibinfo {year} {2022})}\BibitemShut {NoStop}%
\bibitem [{\citenamefont {Kazakov}\ \emph {et~al.}(2021)\citenamefont {Kazakov}, \citenamefont {Opa\v{c}ak}, \citenamefont {Beiser}, \citenamefont {Belyanin}, \citenamefont {Schwarz}, \citenamefont {Piccardo},\ and\ \citenamefont {Capasso}}]{Kazakov21}%
  \BibitemOpen
  \bibfield  {author} {\bibinfo {author} {\bibfnamefont {D.}~\bibnamefont {Kazakov}}, \bibinfo {author} {\bibfnamefont {N.}~\bibnamefont {Opa\v{c}ak}}, \bibinfo {author} {\bibfnamefont {M.}~\bibnamefont {Beiser}}, \bibinfo {author} {\bibfnamefont {A.}~\bibnamefont {Belyanin}}, \bibinfo {author} {\bibfnamefont {B.}~\bibnamefont {Schwarz}}, \bibinfo {author} {\bibfnamefont {M.}~\bibnamefont {Piccardo}}, \ and\ \bibinfo {author} {\bibfnamefont {F.}~\bibnamefont {Capasso}},\ }\bibfield  {title} {\enquote {\bibinfo {title} {Defect-engineered ring laser harmonic frequency combs},}\ }\href {\doibase 10.1364/OPTICA.430896} {\bibfield  {journal} {\bibinfo  {journal} {Optica}\ }\textbf {\bibinfo {volume} {8}},\ \bibinfo {pages} {1277--1280} (\bibinfo {year} {2021})}\BibitemShut {NoStop}%
\bibitem [{\citenamefont {Hillbrand}\ \emph {et~al.}(2018)\citenamefont {Hillbrand}, \citenamefont {Jouy}, \citenamefont {Beck},\ and\ \citenamefont {Faist}}]{Hillbrand18}%
  \BibitemOpen
  \bibfield  {author} {\bibinfo {author} {\bibfnamefont {J.}~\bibnamefont {Hillbrand}}, \bibinfo {author} {\bibfnamefont {P.}~\bibnamefont {Jouy}}, \bibinfo {author} {\bibfnamefont {M.}~\bibnamefont {Beck}}, \ and\ \bibinfo {author} {\bibfnamefont {J.}~\bibnamefont {Faist}},\ }\bibfield  {title} {\enquote {\bibinfo {title} {Tunable dispersion compensation of quantum cascade laser frequency combs},}\ }\href {\doibase 10.1364/OL.43.001746} {\bibfield  {journal} {\bibinfo  {journal} {Opt. Lett.}\ }\textbf {\bibinfo {volume} {43}},\ \bibinfo {pages} {1746--1749} (\bibinfo {year} {2018})}\BibitemShut {NoStop}%
\bibitem [{\citenamefont {Lang}\ and\ \citenamefont {Kobayashi}(1980)}]{LangKobayashi}%
  \BibitemOpen
  \bibfield  {author} {\bibinfo {author} {\bibfnamefont {R.}~\bibnamefont {Lang}}\ and\ \bibinfo {author} {\bibfnamefont {K.}~\bibnamefont {Kobayashi}},\ }\bibfield  {title} {\enquote {\bibinfo {title} {External optical feedback effects on semiconductor injection laser properties},}\ }\href {\doibase 10.1109/JQE.1980.1070479} {\bibfield  {journal} {\bibinfo  {journal} {IEEE Journal of Quantum Electronics}\ }\textbf {\bibinfo {volume} {16}},\ \bibinfo {pages} {347--355} (\bibinfo {year} {1980})}\BibitemShut {NoStop}%
\bibitem [{\citenamefont {Qi}\ \emph {et~al.}(2021{\natexlab{a}})\citenamefont {Qi}, \citenamefont {Bertling}, \citenamefont {Taimre}, \citenamefont {Agnew}, \citenamefont {Lim}, \citenamefont {Gillespie}, \citenamefont {Robinson}, \citenamefont {Br\"unig}, \citenamefont {Demi\ifmmode~\acute{c}\else \'{c}\fi{}}, \citenamefont {Dean}, \citenamefont {Li}, \citenamefont {Linfield}, \citenamefont {Davies}, \citenamefont {Indjin},\ and\ \citenamefont {Raki\ifmmode~\acute{c}\else \'{c}\fi{}}}]{Tina1}%
  \BibitemOpen
  \bibfield  {author} {\bibinfo {author} {\bibfnamefont {X.}~\bibnamefont {Qi}}, \bibinfo {author} {\bibfnamefont {K.}~\bibnamefont {Bertling}}, \bibinfo {author} {\bibfnamefont {T.}~\bibnamefont {Taimre}}, \bibinfo {author} {\bibfnamefont {G.}~\bibnamefont {Agnew}}, \bibinfo {author} {\bibfnamefont {Y.~L.}\ \bibnamefont {Lim}}, \bibinfo {author} {\bibfnamefont {T.}~\bibnamefont {Gillespie}}, \bibinfo {author} {\bibfnamefont {A.}~\bibnamefont {Robinson}}, \bibinfo {author} {\bibfnamefont {M.}~\bibnamefont {Br\"unig}}, \bibinfo {author} {\bibfnamefont {A.}~\bibnamefont {Demi\ifmmode~\acute{c}\else \'{c}\fi{}}}, \bibinfo {author} {\bibfnamefont {P.}~\bibnamefont {Dean}}, \bibinfo {author} {\bibfnamefont {L.~H.}\ \bibnamefont {Li}}, \bibinfo {author} {\bibfnamefont {E.~H.}\ \bibnamefont {Linfield}}, \bibinfo {author} {\bibfnamefont {A.~G.}\ \bibnamefont {Davies}}, \bibinfo {author} {\bibfnamefont {D.}~\bibnamefont {Indjin}}, \ and\ \bibinfo {author} {\bibfnamefont {A.~D.}\ \bibnamefont
  {Raki\ifmmode~\acute{c}\else \'{c}\fi{}}},\ }\bibfield  {title} {\enquote {\bibinfo {title} {Observation of optical feedback dynamics in single-mode terahertz quantum cascade lasers: Transient instabilities},}\ }\href {\doibase 10.1103/PhysRevA.103.033504} {\bibfield  {journal} {\bibinfo  {journal} {Phys. Rev. A}\ }\textbf {\bibinfo {volume} {103}},\ \bibinfo {pages} {033504} (\bibinfo {year} {2021}{\natexlab{a}})}\BibitemShut {NoStop}%
\bibitem [{\citenamefont {Qi}\ \emph {et~al.}(2021{\natexlab{b}})\citenamefont {Qi}, \citenamefont {Bertling}, \citenamefont {Taimre}, \citenamefont {Agnew}, \citenamefont {Lim}, \citenamefont {Gillespie}, \citenamefont {Demi\'{c}}, \citenamefont {Dean}, \citenamefont {Li}, \citenamefont {Linfield}, \citenamefont {Davies}, \citenamefont {Indjin},\ and\ \citenamefont {Raki\'{c}}}]{Tina2}%
  \BibitemOpen
  \bibfield  {author} {\bibinfo {author} {\bibfnamefont {X.}~\bibnamefont {Qi}}, \bibinfo {author} {\bibfnamefont {K.}~\bibnamefont {Bertling}}, \bibinfo {author} {\bibfnamefont {T.}~\bibnamefont {Taimre}}, \bibinfo {author} {\bibfnamefont {G.}~\bibnamefont {Agnew}}, \bibinfo {author} {\bibfnamefont {Y.~L.}\ \bibnamefont {Lim}}, \bibinfo {author} {\bibfnamefont {T.}~\bibnamefont {Gillespie}}, \bibinfo {author} {\bibfnamefont {A.}~\bibnamefont {Demi\'{c}}}, \bibinfo {author} {\bibfnamefont {P.}~\bibnamefont {Dean}}, \bibinfo {author} {\bibfnamefont {L.~H.}\ \bibnamefont {Li}}, \bibinfo {author} {\bibfnamefont {E.~H.}\ \bibnamefont {Linfield}}, \bibinfo {author} {\bibfnamefont {A.~G.}\ \bibnamefont {Davies}}, \bibinfo {author} {\bibfnamefont {D.}~\bibnamefont {Indjin}}, \ and\ \bibinfo {author} {\bibfnamefont {A.~D.}\ \bibnamefont {Raki\'{c}}},\ }\bibfield  {title} {\enquote {\bibinfo {title} {Terahertz quantum cascade laser under optical feedback: effects of laser self-pulsations on self-mixing signals},}\
  }\href {\doibase 10.1364/OE.437861} {\bibfield  {journal} {\bibinfo  {journal} {Opt. Express}\ }\textbf {\bibinfo {volume} {29}},\ \bibinfo {pages} {39885--39894} (\bibinfo {year} {2021}{\natexlab{b}})}\BibitemShut {NoStop}%
\bibitem [{\citenamefont {Silvestri}\ \emph {et~al.}(2020)\citenamefont {Silvestri}, \citenamefont {Columbo}, \citenamefont {Brambilla},\ and\ \citenamefont {Gioannini}}]{Silvestri20}%
  \BibitemOpen
  \bibfield  {author} {\bibinfo {author} {\bibfnamefont {C.}~\bibnamefont {Silvestri}}, \bibinfo {author} {\bibfnamefont {L.~L.}\ \bibnamefont {Columbo}}, \bibinfo {author} {\bibfnamefont {M.}~\bibnamefont {Brambilla}}, \ and\ \bibinfo {author} {\bibfnamefont {M.}~\bibnamefont {Gioannini}},\ }\bibfield  {title} {\enquote {\bibinfo {title} {Coherent multi-mode dynamics in a quantum cascade laser: amplitude- and frequency-modulated optical frequency combs},}\ }\href {\doibase 10.1364/OE.396481} {\bibfield  {journal} {\bibinfo  {journal} {Opt. Express}\ }\textbf {\bibinfo {volume} {28}},\ \bibinfo {pages} {23846--23861} (\bibinfo {year} {2020})}\BibitemShut {NoStop}%
\bibitem [{\citenamefont {Silvestri}\ \emph {et~al.}(2021)\citenamefont {Silvestri}, \citenamefont {Columbo}, \citenamefont {Brambilla},\ and\ \citenamefont {Gioannini}}]{SilvestriCLEO}%
  \BibitemOpen
  \bibfield  {author} {\bibinfo {author} {\bibfnamefont {C.}~\bibnamefont {Silvestri}}, \bibinfo {author} {\bibfnamefont {L.~L.}\ \bibnamefont {Columbo}}, \bibinfo {author} {\bibfnamefont {M.}~\bibnamefont {Brambilla}}, \ and\ \bibinfo {author} {\bibfnamefont {M.}~\bibnamefont {Gioannini}},\ }\bibfield  {title} {\enquote {\bibinfo {title} {Dynamics of optical frequency combs in ring and {F}abry-{P}erot {Q}uantum {C}ascade {L}asers},}\ }in\ \href {\doibase 10.1109/CLEO/Europe-EQEC52157.2021.9541633} {\emph {\bibinfo {booktitle} {2021 Conference on Lasers and Electro-Optics Europe and European Quantum Electronics Conference (CLEO/Europe-EQEC)}}}\ (\bibinfo {year} {2021})\ pp.\ \bibinfo {pages} {1--1}\BibitemShut {NoStop}%
\bibitem [{\citenamefont {Silvestri}\ \emph {et~al.}(2022)\citenamefont {Silvestri}, \citenamefont {Qi}, \citenamefont {Taimre},\ and\ \citenamefont {Raki\ifmmode~\acute{c}\else \'{c}\fi{}}}]{Silvestri22}%
  \BibitemOpen
  \bibfield  {author} {\bibinfo {author} {\bibfnamefont {C.}~\bibnamefont {Silvestri}}, \bibinfo {author} {\bibfnamefont {X.}~\bibnamefont {Qi}}, \bibinfo {author} {\bibfnamefont {T.}~\bibnamefont {Taimre}}, \ and\ \bibinfo {author} {\bibfnamefont {A.~D.}\ \bibnamefont {Raki\ifmmode~\acute{c}\else \'{c}\fi{}}},\ }\bibfield  {title} {\enquote {\bibinfo {title} {Multimode dynamics of terahertz quantum cascade lasers: Spontaneous and actively induced generation of dense and harmonic coherent regimes},}\ }\href {\doibase 10.1103/PhysRevA.106.053526} {\bibfield  {journal} {\bibinfo  {journal} {Phys. Rev. A}\ }\textbf {\bibinfo {volume} {106}},\ \bibinfo {pages} {053526} (\bibinfo {year} {2022})}\BibitemShut {NoStop}%
\bibitem [{\citenamefont {Li}\ \emph {et~al.}(2015)\citenamefont {Li}, \citenamefont {Laffaille}, \citenamefont {Gacemi}, \citenamefont {Apfel}, \citenamefont {Sirtori}, \citenamefont {Leonardon}, \citenamefont {Santarelli}, \citenamefont {R\"{o}sch}, \citenamefont {Scalari}, \citenamefont {Beck}, \citenamefont {Faist}, \citenamefont {H\"{a}nsel}, \citenamefont {Holzwarth},\ and\ \citenamefont {Barbieri}}]{Li15}%
  \BibitemOpen
  \bibfield  {author} {\bibinfo {author} {\bibfnamefont {H.}~\bibnamefont {Li}}, \bibinfo {author} {\bibfnamefont {P.}~\bibnamefont {Laffaille}}, \bibinfo {author} {\bibfnamefont {D.}~\bibnamefont {Gacemi}}, \bibinfo {author} {\bibfnamefont {M.}~\bibnamefont {Apfel}}, \bibinfo {author} {\bibfnamefont {C.}~\bibnamefont {Sirtori}}, \bibinfo {author} {\bibfnamefont {J.}~\bibnamefont {Leonardon}}, \bibinfo {author} {\bibfnamefont {G.}~\bibnamefont {Santarelli}}, \bibinfo {author} {\bibfnamefont {M.}~\bibnamefont {R\"{o}sch}}, \bibinfo {author} {\bibfnamefont {G.}~\bibnamefont {Scalari}}, \bibinfo {author} {\bibfnamefont {M.}~\bibnamefont {Beck}}, \bibinfo {author} {\bibfnamefont {J.}~\bibnamefont {Faist}}, \bibinfo {author} {\bibfnamefont {W.}~\bibnamefont {H\"{a}nsel}}, \bibinfo {author} {\bibfnamefont {R.}~\bibnamefont {Holzwarth}}, \ and\ \bibinfo {author} {\bibfnamefont {S.}~\bibnamefont {Barbieri}},\ }\bibfield  {title} {\enquote {\bibinfo {title} {Dynamics of ultra-broadband terahertz quantum cascade lasers
  for comb operation},}\ }\href {\doibase 10.1364/OE.23.033270} {\bibfield  {journal} {\bibinfo  {journal} {Opt. Express}\ }\textbf {\bibinfo {volume} {23}},\ \bibinfo {pages} {33270--33294} (\bibinfo {year} {2015})}\BibitemShut {NoStop}%
\bibitem [{\citenamefont {Rimoldi}\ \emph {et~al.}(2022)\citenamefont {Rimoldi}, \citenamefont {Columbo}, \citenamefont {Bovington}, \citenamefont {Romero-Garc\'{i}a},\ and\ \citenamefont {Gioannini}}]{Rimoldi22}%
  \BibitemOpen
  \bibfield  {author} {\bibinfo {author} {\bibfnamefont {C.}~\bibnamefont {Rimoldi}}, \bibinfo {author} {\bibfnamefont {L.~L.}\ \bibnamefont {Columbo}}, \bibinfo {author} {\bibfnamefont {J.}~\bibnamefont {Bovington}}, \bibinfo {author} {\bibfnamefont {S.}~\bibnamefont {Romero-Garc\'{i}a}}, \ and\ \bibinfo {author} {\bibfnamefont {M.}~\bibnamefont {Gioannini}},\ }\bibfield  {title} {\enquote {\bibinfo {title} {Damping of relaxation oscillations, photon-photon resonance, and tolerance to external optical feedback of iii-v/sin hybrid lasers with a dispersive narrow band mirror},}\ }\href {\doibase 10.1364/OE.452155} {\bibfield  {journal} {\bibinfo  {journal} {Opt. Express}\ }\textbf {\bibinfo {volume} {30}},\ \bibinfo {pages} {11090--11109} (\bibinfo {year} {2022})}\BibitemShut {NoStop}%
\bibitem [{\citenamefont {Bacon}\ \emph {et~al.}(2016)\citenamefont {Bacon}, \citenamefont {Freeman}, \citenamefont {Mohandas}, \citenamefont {Li}, \citenamefont {Linfield}, \citenamefont {Davies},\ and\ \citenamefont {Dean}}]{THz_taue_1}%
  \BibitemOpen
  \bibfield  {author} {\bibinfo {author} {\bibfnamefont {D.~R.}\ \bibnamefont {Bacon}}, \bibinfo {author} {\bibfnamefont {J.~R.}\ \bibnamefont {Freeman}}, \bibinfo {author} {\bibfnamefont {R.~A.}\ \bibnamefont {Mohandas}}, \bibinfo {author} {\bibfnamefont {L.}~\bibnamefont {Li}}, \bibinfo {author} {\bibfnamefont {E.~H.}\ \bibnamefont {Linfield}}, \bibinfo {author} {\bibfnamefont {A.~G.}\ \bibnamefont {Davies}}, \ and\ \bibinfo {author} {\bibfnamefont {P.}~\bibnamefont {Dean}},\ }\bibfield  {title} {\enquote {\bibinfo {title} {Gain recovery time in a terahertz quantum cascade laser},}\ }\href {\doibase https://doi.org/10.1002/advs.202200410} {\bibfield  {journal} {\bibinfo  {journal} {Applied Physics Letters}\ }\textbf {\bibinfo {volume} {108}},\ \bibinfo {pages} {081104} (\bibinfo {year} {2016})}\BibitemShut {NoStop}%
\bibitem [{\citenamefont {Leng~Lim}\ \emph {et~al.}(2011)\citenamefont {Leng~Lim}, \citenamefont {Dean}, \citenamefont {Nikolić}, \citenamefont {Kliese}, \citenamefont {Khanna}, \citenamefont {Lachab}, \citenamefont {Valavanis}, \citenamefont {Indjin}, \citenamefont {Ikonić}, \citenamefont {Harrison}, \citenamefont {Linfield}, \citenamefont {Giles~Davies}, \citenamefont {Wilson},\ and\ \citenamefont {Rakić}}]{alphaneg}%
  \BibitemOpen
  \bibfield  {author} {\bibinfo {author} {\bibfnamefont {Y.}~\bibnamefont {Leng~Lim}}, \bibinfo {author} {\bibfnamefont {P.}~\bibnamefont {Dean}}, \bibinfo {author} {\bibfnamefont {M.}~\bibnamefont {Nikolić}}, \bibinfo {author} {\bibfnamefont {R.}~\bibnamefont {Kliese}}, \bibinfo {author} {\bibfnamefont {S.~P.}\ \bibnamefont {Khanna}}, \bibinfo {author} {\bibfnamefont {M.}~\bibnamefont {Lachab}}, \bibinfo {author} {\bibfnamefont {A.}~\bibnamefont {Valavanis}}, \bibinfo {author} {\bibfnamefont {D.}~\bibnamefont {Indjin}}, \bibinfo {author} {\bibfnamefont {Z.}~\bibnamefont {Ikonić}}, \bibinfo {author} {\bibfnamefont {P.}~\bibnamefont {Harrison}}, \bibinfo {author} {\bibfnamefont {E.~H.}\ \bibnamefont {Linfield}}, \bibinfo {author} {\bibfnamefont {A.}~\bibnamefont {Giles~Davies}}, \bibinfo {author} {\bibfnamefont {S.~J.}\ \bibnamefont {Wilson}}, \ and\ \bibinfo {author} {\bibfnamefont {A.~D.}\ \bibnamefont {Rakić}},\ }\bibfield  {title} {\enquote {\bibinfo {title} {{Demonstration of a self-mixing displacement
  sensor based on terahertz quantum cascade lasers}},}\ }\href {\doibase https://doi.org/10.1063/1.3629991} {\bibfield  {journal} {\bibinfo  {journal} {Applied Physics Letters}\ }\textbf {\bibinfo {volume} {99}} (\bibinfo {year} {2011}),\ https://doi.org/10.1063/1.3629991}\BibitemShut {NoStop}%
\bibitem [{\citenamefont {Green}\ \emph {et~al.}(2008)\citenamefont {Green}, \citenamefont {Xu}, \citenamefont {Mahler}, \citenamefont {Tredicucci}, \citenamefont {Beltram}, \citenamefont {Giuliani}, \citenamefont {Beere},\ and\ \citenamefont {Ritchie}}]{negalpha1}%
  \BibitemOpen
  \bibfield  {author} {\bibinfo {author} {\bibfnamefont {R.~P.}\ \bibnamefont {Green}}, \bibinfo {author} {\bibfnamefont {J.-H.}\ \bibnamefont {Xu}}, \bibinfo {author} {\bibfnamefont {L.}~\bibnamefont {Mahler}}, \bibinfo {author} {\bibfnamefont {A.}~\bibnamefont {Tredicucci}}, \bibinfo {author} {\bibfnamefont {F.}~\bibnamefont {Beltram}}, \bibinfo {author} {\bibfnamefont {G.}~\bibnamefont {Giuliani}}, \bibinfo {author} {\bibfnamefont {H.~E.}\ \bibnamefont {Beere}}, \ and\ \bibinfo {author} {\bibfnamefont {D.~A.}\ \bibnamefont {Ritchie}},\ }\bibfield  {title} {\enquote {\bibinfo {title} {{Linewidth enhancement factor of terahertz quantum cascade lasers}},}\ }\href {\doibase 10.1063/1.2883950} {\bibfield  {journal} {\bibinfo  {journal} {Applied Physics Letters}\ }\textbf {\bibinfo {volume} {92}},\ \bibinfo {pages} {071106} (\bibinfo {year} {2008})},\ \Eprint {http://arxiv.org/abs/https://pubs.aip.org/aip/apl/article-pdf/doi/10.1063/1.2883950/14080565/071106\_1\_online.pdf}
  {https://pubs.aip.org/aip/apl/article-pdf/doi/10.1063/1.2883950/14080565/071106\_1\_online.pdf} \BibitemShut {NoStop}%
\bibitem [{\citenamefont {von Staden}\ \emph {et~al.}(2006)\citenamefont {von Staden}, \citenamefont {Gensty}, \citenamefont {Els\"{a}{\ss}er}, \citenamefont {Giuliani},\ and\ \citenamefont {Mann}}]{negalpha2}%
  \BibitemOpen
  \bibfield  {author} {\bibinfo {author} {\bibfnamefont {J.}~\bibnamefont {von Staden}}, \bibinfo {author} {\bibfnamefont {T.}~\bibnamefont {Gensty}}, \bibinfo {author} {\bibfnamefont {W.}~\bibnamefont {Els\"{a}{\ss}er}}, \bibinfo {author} {\bibfnamefont {G.}~\bibnamefont {Giuliani}}, \ and\ \bibinfo {author} {\bibfnamefont {C.}~\bibnamefont {Mann}},\ }\bibfield  {title} {\enquote {\bibinfo {title} {Measurements of the $\alpha$ factor of a distributed-feedback quantum cascade laser by an optical feedback self-mixing technique},}\ }\href {\doibase 10.1364/OL.31.002574} {\bibfield  {journal} {\bibinfo  {journal} {Opt. Lett.}\ }\textbf {\bibinfo {volume} {31}},\ \bibinfo {pages} {2574--2576} (\bibinfo {year} {2006})}\BibitemShut {NoStop}%
\bibitem [{\citenamefont {Piccardo}\ \emph {et~al.}(2018{\natexlab{b}})\citenamefont {Piccardo}, \citenamefont {Chevalier}, \citenamefont {Mansuripur}, \citenamefont {Kazakov}, \citenamefont {Wang}, \citenamefont {Rubin}, \citenamefont {Meadowcroft}, \citenamefont {Belyanin},\ and\ \citenamefont {Capasso}}]{PiccardoHFCOptex}%
  \BibitemOpen
  \bibfield  {author} {\bibinfo {author} {\bibfnamefont {M.}~\bibnamefont {Piccardo}}, \bibinfo {author} {\bibfnamefont {P.}~\bibnamefont {Chevalier}}, \bibinfo {author} {\bibfnamefont {T.~S.}\ \bibnamefont {Mansuripur}}, \bibinfo {author} {\bibfnamefont {D.}~\bibnamefont {Kazakov}}, \bibinfo {author} {\bibfnamefont {Y.}~\bibnamefont {Wang}}, \bibinfo {author} {\bibfnamefont {N.~A.}\ \bibnamefont {Rubin}}, \bibinfo {author} {\bibfnamefont {L.}~\bibnamefont {Meadowcroft}}, \bibinfo {author} {\bibfnamefont {A.}~\bibnamefont {Belyanin}}, \ and\ \bibinfo {author} {\bibfnamefont {F.}~\bibnamefont {Capasso}},\ }\bibfield  {title} {\enquote {\bibinfo {title} {The harmonic state of quantum cascade lasers: origin, control, and prospective applications},}\ }\href {\doibase https://doi.org/10.1364/OE.26.009464} {\bibfield  {journal} {\bibinfo  {journal} {Opt. Express}\ }\textbf {\bibinfo {volume} {26}},\ \bibinfo {pages} {9464--9483} (\bibinfo {year} {2018}{\natexlab{b}})}\BibitemShut {NoStop}%
\bibitem [{\citenamefont {Columbo}\ and\ \citenamefont {Brambilla}(2014)}]{columbo2014}%
  \BibitemOpen
  \bibfield  {author} {\bibinfo {author} {\bibfnamefont {L.~L.}\ \bibnamefont {Columbo}}\ and\ \bibinfo {author} {\bibfnamefont {M.}~\bibnamefont {Brambilla}},\ }\bibfield  {title} {\enquote {\bibinfo {title} {Multimode regimes in quantum cascade lasers with optical feedback},}\ }\href {\doibase 10.1364/OE.22.010105} {\bibfield  {journal} {\bibinfo  {journal} {Opt. Express}\ }\textbf {\bibinfo {volume} {22}},\ \bibinfo {pages} {10105--10118} (\bibinfo {year} {2014})}\BibitemShut {NoStop}%
\end{thebibliography}%

\end{document}


\renewcommand{\theequation}{S.\arabic{equation}}
\renewcommand{\thefigure}{S.\arabic{figure}}
\renewcommand{\thesection}{S.\arabic{section}}
\preprint{AIP/123-QED}


\title{Frequency combs induced by optical feedback and harmonic order tunability in quantum cascade lasers: supplementary materials}
\author{Carlo Silvestri}
\affiliation{School of Electrical Engineering and Computer Science, The University of Queensland, Brisbane, QLD 4072, Australia
}

\author{Xiaoqiong Qi}%
\affiliation{School of Electrical Engineering and Computer Science, The University of Queensland, Brisbane, QLD 4072, Australia
}
\author{Thomas Taimre}
\affiliation{School of Mathematics and Physics, The University of Queensland, Brisbane, QLD 4072, Australia
}

\author{Aleksandar D. Raki\'c}
\affiliation{School of Electrical Engineering and Computer Science, The University of Queensland, Brisbane, QLD 4072, Australia
}

\date{\today}

\maketitle

\section{Effective semiconductor Maxwell-Bloch equations with optical feedback}
We study the quantum cascade laser (QCL) dynamics by using a full set of effective semiconductor Maxwell--Bloch equations (ESMBEs) for the Fabry--Perot (FP) configuration.\cite{Silvestri20,Silvestri22, SilvestriCLEO} This model is based on a phenomenological expression for the optical susceptibility of a QCL active medium, and it encompasses the main properties of semiconductor materials such as a non null $\alpha$ factor, a dependence of the susceptibility from the density of carriers, asymmetric gain and refractive index spectra, and nonlinearities at each order, such as Kerr effect and four-wave mixing. In the case of FP cavity it also accounts for SHB through the inclusion of a carrier grating. The ESMBEs read:
\begin{widetext}
\begin{eqnarray}
\frac{\partial E^+}{\partial z}+ \frac{1}{v}\frac{\partial E^+}{\partial t} &=& -\frac{\alpha_\mathrm{L}}{2}E^++g P_0^+ ,\label{el+}\\
-\frac{\partial E^-}{\partial z}+ \frac{1}{v}\frac{\partial E^-}{\partial t} &=& -\frac{\alpha_\mathrm{L}}{2}E^-+g P_0^- ,\label{el-}\\
\frac{\partial P_0^+}{\partial t}&=&\mathrm{\pi\delta_{hom}}(1+i\alpha)\left[-P_0^++if_0\varepsilon_0\varepsilon_\mathrm{b}\left(1+i\alpha\right)\left(N_0E^{+}+N_1^+E^-\right) \right], \label{P+}\\
\frac{\partial P_0^-}{\partial t}&=&\mathrm{\pi\delta_{hom}}(1+i\alpha)\left[-P_0^-+if_0\varepsilon_0\varepsilon_\mathrm{b}\left(1+i\alpha\right)\left(N_0E^{-}+N_1^-E^+\right) \right], \label{P-}\\
\frac{\partial N_0}{\partial t}&=&\frac{I}{eV}-\frac{N_0}{\tau_\mathrm{e}}+\frac{i}{4\hbar}\left[E^{+*}P_0^++E^{-*}P_0^--E^{+}P_0^{+*}-E^{-}P_0^{-*}\right], \label{N0}\\
\frac{\partial N_1^+}{\partial t}&=&-\frac{N_1^+}{\tau_\mathrm{e}}+\frac{i}{4\hbar}\left[E^{-*}P_0^+-E^{+}P_0^{-*}\right]. \label{N1}
\end{eqnarray}
\end{widetext}
where $E^+(z,t)$, $E^-(z,t)$ are the forward and backward electric fields, $P_0^+$, $P_0^-$ are the forward and backward polarization terms, $N_0$ is the zero-order density of carriers, and $N_1^+$, $N_1^-$ are the first-order terms of the density of carriers, reproducing the carrier grating due to SHB; $v$ is the group velocity, $\alpha_\mathrm{L}$ is the loss coefficient, $\alpha$ is the linewidth enhancement factor, $\delta_\mathrm{{hom}}$ is the homogeneous part of the gain bandwidth, $f_0$ is the differential gain, $\varepsilon_0$ is the dielectric permittivity of the vacuum, $\varepsilon_\mathrm{b}$ is the relative dielectric constant of the QCL active medium, $I$ is the driving current of the laser, $V$ is the volume of the active region, $\tau_\mathrm{e}$ is the carrier lifetime, and the coefficient $g$ is given by:
\begin{equation}
g=\frac{-i\omega_0N_\mathrm{p}\Gamma_\mathrm{c}}{2\varepsilon_0n_\mathrm{r}c} \, , \label{g}
\end{equation}
where $N_\mathrm{p}$ is the number of stages of structure, $\omega_0$ is the cold cavity angular frequency closest to the gain peak and it is used as a reference frequency, $\Gamma_\mathrm{c}$ is the optical confinement factor, $c$ is the speed of light in the vacuum, and $n_\mathrm{r}$ is the effective background refractive index of the medium. We remark that $\delta_\mathrm{hom}$ is related to the polarization dephasing time $\tau_\mathrm{d}$ through the equation $\delta_{\mathrm{hom}}=\frac{1}{\pi \tau_\mathrm{d}}$.\cite{Silvestri20}\\
We want to include the effect of the optical feedback in the boundary conditions of the model. Therefore, we start by adding a term $E_\mathrm{fb}(t)$ to the free-running boundary conditions, corresponding to an external target placed at distance $L_\mathrm{ext}$ from right facet of the QCL:
\begin{eqnarray}
E^-(L, t)&=&\sqrt{R}E^+(L,t)+E_\mathrm{fb}(t),\label{bc2fb1}\\
E^+(0, t)&=&\sqrt{R}E^-(0,t),\label{bc2fb2}
\end{eqnarray}\\
where $L$ is the QCL cavity length and $R$ is the reflectivity of each facet of the QCL. We want to determine $E_\mathrm{fb}(t)$. In the frequency domain we can write the reflection coefficient due to the external feedback in our configuration as:
\begin{eqnarray}
r(\omega)=r_\mathrm{ext}t_\mathrm{L}^2e^{-i\Phi_\mathrm{ext}}e^{-i(\omega-\omega_0)\tau_\mathrm{ext}}\label{r_omega}
\end{eqnarray}
where $\omega$ is the angular frequency, $t_\mathrm{L}$ is the field transmissivity for both outgoing and incoming field, $\Phi_\mathrm{ext}=2L_\mathrm{ext}\omega_0/c$, $\tau_\mathrm{ext}=2L_\mathrm{ext}/c$, $r_\mathrm{ext}$ is the frequency-independent reflectivity of the external target. If we calculate the Fourier transform of the forward field $E^+(L,t)$, we multiply it by $r(\omega)$ given by  Eq.~(\ref{r_omega}), and then we anti-transform the obtained product, we retrieve the following expression for the feedback term.\cite{Rimoldi22}
\begin{eqnarray}
E_\mathrm{fb}(t)=r_\mathrm{ext}t_\mathrm{L}^2E^+\left(L, t-\tau_\mathrm{ext}\right)\label{efb2}
\end{eqnarray}
Since Eqs.~(\ref{el+})--(\ref{N1}) have been derived by assuming the frequencies of the multimode fields are referenced with respect to the central frequency $\omega_0$ (see ref.\cite{Silvestri20}), we write Eq.~(\ref{efb2}) consistently with this hypothesis:
\begin{eqnarray}
E_\mathrm{fb}(t)=r_\mathrm{ext}t_\mathrm{L}^2E^+\left(L, t-\tau_\mathrm{ext}\right)e^{-i\omega_0\tau_\mathrm{ext}}\label{efb2}
\end{eqnarray}
If we also take into account the losses in the external cavity $\epsilon_\mathrm{L}$, and the ones due to the reinjection $\epsilon_\mathrm{S}$, we obtain the boundary conditions Eqs.~(1)--(2) presented in the main manuscript.\\

\section{Numerical study: procedure and methods}
\begin{table}[h]
\begin{center}
\begin{tabular}{ |l|l| } 
\hline
 Length of the laser cavity & $L=2$ mm  \\ 
 Effective refractive index & $n=3.6$  \\ 
 Confinement factor & $\Gamma_c=0.13$ \\
 Mirror reflectivity & $R=0.3$ \\
 Differential gain & $f_0= 7\times10^{-5} \mu\mathrm{m}^3$  \\
 Volume of the active region & $V=3.6 \times10^{6} \mu\mathrm{m}^3$  \\
 Number of stages & $N_p=90$  \\
 Homogeneous gain bandwidth & $\delta_\mathrm{hom}=0.32$ THz  \\
 Waveguide loss & $\alpha_\mathrm{L}=3.8~\mathrm{cm^{-1}}$   \\
 Central emission frequency & $\nu_0=3$ THz  \\
 Linewidth enhancement factor & varies  \\
 Carrier lifetime & varies  \\
 \hline
\end{tabular}
\caption{\label{Table1} ESMBEs parameters. \cite{PiccardoReview,SilvestriReview,alphaneg,Silvestri22,Faist_2016, Li22, Li15,Vitiello2021}}
\end{center}
\end{table}
The values of the ESMBEs parameters exploited in this article are presented in Table I, and they are typical values for a THz QCL.\cite{SilvestriReview,PiccardoReview} They have been used to study the free-running case in ref.~\cite{Silvestri22}, where two comb regions characterized by fundamental and second order harmonic combs were found.\\
\begin{figure}[t] 
\centering
\includegraphics[width=0.4\textwidth]{FiguresSupplementary/sim_standard.png}
\caption{Example of a performed simulation. For the first $40~$ns, the QCL is configured in free-running operation: after a time interval of about $33$ ns in which no power is emitted (i), the QCL starts to lase and reaches a stable single mode emission (ii); then, at $t=40$~ns, the optical feedback is switched on, and the QCL transits on a different dynamical regime (iii).}
\label{simstand}
\end{figure}
The numerical study which leads to Fig.~2 presented in the main manuscript, is conducted according to the following points. The laser is initially configured in free running operation, i.e. we solve the ESMBEs by adopting the boundary conditions with $\epsilon=0$, and the value of the bias current is chosen in order to have single mode emission; in particular, we set $I=1.08I_\mathrm{thr}$. Once the QCL output reaches a stable single mode emission, at $t=40$ ns the external target is introduced (Fig.~\ref{simstand}) by integrating the ESMBEs with the boundary conditions Eqs.~(1)--(2), for fixed values of $L_{\mathrm{ext}}$ and $\epsilon$. Each simulation has total duration of $1\mu s$ and is performed for a pair ($\epsilon$, $L_{\mathrm{ext}}$); the set of values of $\epsilon$ are chosen between 0.01 and 0.09 with step 0.01, and between 0.1 and 1 with step 0.1; the values of $L_{\mathrm{ext}}$ are selected in four regions with a different step for each region: region 1 corresponds to the short cavity regime, and includes the values between 1~mm and 9~mm, with step 1~mm; region 2 includes the values between 1~cm and 9~cm, with step 1~cm; region 3 corresponds to $L_\mathrm{ext}$ between 10~cm and 90~cm with step 10~cm; in region 4 we consider the long cavity regime, with the values 1~m, 2~m, and 3~m. This choice allows us to study the laser dynamics in different ranges of  $L_\mathrm{ext}$ with a limited number of simulations, providing a general overview of the laser behaviour under optical feedback.\\
The set of ESMBEs~(\ref{el+})-(\ref{N1}) with the boundary conditions (\ref{bc2fb1})-(\ref{bc2fb2}) have been integrated using an optimized finite difference algorithm discretizing in both time and space, described in detail in the Appendix of Ref.~\cite{Bardella17}. The time step used in the simulations is $dt=20~$fs, which corresponds to a space step $dz=1.8~\mu$m. This leads to a duration for a single typical $1~\mu$s long simulation of 5 hours on our Intel$\copyright$ Xeon$\copyright$ W-2145 CPU 3.70 GHz processor, with 64 GB RAM. However, we specify that the duration of the transients preceding the attainment of the steady state condition varies depending on the feedback parameters. This implies that simulations have slightly different durations from each other, and the values of 1 $\mu$s and 5 hours are therefore indicative (average) values.

We would like to highlight that the study described in Section IV of the main manuscript follows the same procedure, with one difference: the bias current value is set to $I=1.5I_\mathrm{thr}$, which corresponds to a self-starting optical frequency comb regime. Therefore, we first simulate the free-running laser ($\epsilon=0$), allowing it to reach a steady state comb emission. Subsequently, we switch on the feedback and reproduce the map depicted in Fig.~5(a).\\\\
Finally, after solving the ESMBEs for each pair ($\epsilon$, $L_{\mathrm{ext}}$), we classify the obtained dynamical regimes following a rigorous procedure described in the following subsection.
\subsection{Classification of the dynamical regimes}
In order to classify the dynamical regimes, we filter the optical spectrum and we consider the modes within a 20dB power ratio to the spectral maximum. Let us name $N_{20}$ the number of modes included in the $-$20dB bandwidth from the peak in the optical spectrum. If $N_{20}=1$, i.e. only the spectral maximum is included in $-20~\mathrm{dB}$ spectral bandwidth, we classify the regime as a single mode emission.\\
If $N_{20}>1$, we compute the the comb indicators $M_{\sigma_P}$ and $M_{\Delta\Phi}$, previously introduced and exploited in refs.~\cite{Silvestri20, Silvestri22}, which quantify respectively the power and phase noise, and allow us to understand if a regime is locked. Low values of these indicators correspond to low values of amplitude and phase noise, and therefore correspond to a frequency comb regime. We calculate them by filtering each line $q$ of the output optical field in the $-20~\mathrm{dB}$ spectral bandwidth, and we retrieve the dynamics of the power of each line ($P_q(t), q=1,...,N_{20}$) and the intermodal phase difference ($\Delta\Phi_q(t), q=1,...,N_{20}$). Then, we define the comb indicators by using the following equations:
\begin{eqnarray}
M_{\sigma_{P}}&=&\frac{1}{N_{20}}\sum_{q=1}^{N_{10}}{\sigma_{P_q}}\\ M_{\Delta\Phi}&=&\frac{1}{N_{20}}\sum_{q=1}^{N_{20}}{\sigma_{\Delta\Phi_q}},\label{MsigmaMPhi}
\end{eqnarray}
where:
\begin{eqnarray}
\mu_{P_q}&=&\left\langle P_q(t)\right\rangle \\
\mu_{\Delta\Phi_q}&=&\left\langle\Delta\Phi_q(t)\right\rangle,
\end{eqnarray}
\begin{eqnarray}
\sigma_{P_q}&=&\sqrt{\left\langle\left(P_q(t)-\mu_{P_q}\right)^2\right\rangle}\\
\sigma_{\Delta\Phi_q}&=&\sqrt{\left\langle\left(\Delta\Phi_q(t)-\mu_{\Delta\Phi_q}\right)^2\right\rangle}.
\end{eqnarray}
where the symbol $\left\langle\right\rangle$ indicates the time average.\\ A this point, we use the following criterion to classify the simulated multimode states:
\begin{itemize}
    \item We define a simulated regime as an OFC if $M_{\sigma_{P}}<2\cdot10^{-2} \mathrm{mW}$ and $M_{\Delta \Phi}<2\cdot10^{-2} \mathrm{rad}$, analogously to ref.~\cite{Silvestri22}.
    \item If the simulated dynamics does not correspond to an OFC, we examine its optical spectrum. If the optical spectrum consists of a set of frequency bands whose central frequencies are the modes of the laser cavity, and whose spacing is the free spectral range of the external cavity $\mathrm{FSR_{ext}}$, we classify the regime as a mixed state.
    \item If the simulated dynamics does not correspond to an OFC, and also does not satisfy the condition to be a mixed state, we classify it as an irregular regime.
\end{itemize}

\section{Fine tuning of the external cavity length for a feedback-induced optical frequency comb (single mode as initial condition)}\label{finetuningCW}
In this subsection we show that an OFC generated in regime of short cavity starting from a free-running CW, is sensitive to the phase accumulated in the external cavity. We shift external cavity  by a quantity $\Delta L_\mathrm{ext}$ in the same order of the QCL central wavelength $\lambda=100~\mu$m. In particular we consider integer multiples of $\lambda/4$. As example, in Fig.~\ref{ft} we show how the fundamental comb of Fig.~3(d) in the main manuscript evolves if we introduce a shift $\Delta L_\mathrm{ext}=n\lambda/4$, with n$=1,2,3,4$, reporting alternance CW-OFC with period given by $\lambda/2$. We understand, therefore, that a $\pi$ phase accumulated in the external cavity (corresponding to $\lambda/4$ and $3\lambda/4$) leads to a single-mode emission, while a $2\pi$ phase shift does not affect the comb emission. Therefore, the system is phase-sensitive, and this can be exploited for applications such as the detection of microscopic vibrations, or displacements.\\
\begin{figure}
\centering
\includegraphics[width=0.5\textwidth]{FiguresSupplementary/Figura_LC_7mm_36_perc.png}
\caption{Fine tuning of the EC cavity length $L_\mathrm{ext}$ for the OFC of Fig.~3(d) in the main manuscript. Temporal evolution of the power for different values of $L_\mathrm{ext}=L_\mathrm{ext,0}+n\lambda/4$, with n$=1,2,3,4$. $L_\mathrm{ext,0}=7~$mm, $\epsilon^2=36~\%$, $I=1.08I_\mathrm{thr}$,  other parameters as in Table~\ref{Table1}.}
\label{ft}
\end{figure}
Furthermore, we want to investigate how the comb emission is affected by a fine tuning of the external cavity length in the long cavity regime. If we keep $\epsilon^2=36~\%$ as in the case of Fig.~\ref{ft}, but we consider $L_\mathrm{ext,0}=70~$cm, we report a mixed state emission. If also in this case we introduce a shift $\Delta L_\mathrm{ext}=n\lambda/4$, with n$=1,2,3,4$ of the external cavity length, we report a mixed state for all the considered values of $\Delta L_\mathrm{ext}$, as shown in Fig.~\ref{ft2}.
\begin{figure}
\centering
\includegraphics[width=0.5\textwidth]{FiguresSupplementary/Figura_LC_70cm_36_perc.png}
\caption{Fine tuning of the EC cavity length $L_\mathrm{ext}$ in the long cavity regime for a mixed state case. Temporal evolution of the power for different values of $L_\mathrm{ext}=L_\mathrm{ext,0}+n\lambda/4$, with n$=1,2,3,4$. $L_\mathrm{ext,0}=70~$cm, $\epsilon^2=36~\%$, $I=1.08I_\mathrm{thr}$,  other parameters as in Table~\ref{Table1}.}
\label{ft2}
\end{figure}
At this point we want to examine how a feedback-induced comb is affected by a fine tuning of the external cavity length in the long cavity regime. For this reason, we consider $L_\mathrm{ext,0}=30~$cm and $\epsilon^2=1~\%$, which corresponds to a frequency comb emission in the map of Fig. 2(a) in the main manuscript. By repeating the study performed for the two previous cases, we find that the QCL emits a comb for all the considered values of $\Delta L_\mathrm{ext}$, as depicted in Fig.~\ref{ft3}.
\begin{figure}
\centering
\includegraphics[width=0.5\textwidth]{FiguresSupplementary/Figura_LC_30_4_perc.png}
\caption{Fine tuning of the EC cavity length $L_\mathrm{ext}$ in the long cavity regime for a feedback-induced OFC. Temporal evolution of the power for different values of $L_\mathrm{ext}=L_\mathrm{ext,0}+n\lambda/4$, with n$=1,2,3,4$. $L_\mathrm{ext,0}=30~$cm, $\epsilon^2=1~\%$, $I=1.08I_\mathrm{thr}$,  other parameters as in Table~\ref{Table1}.}
\label{ft3}
\end{figure}
We can notice that also in this case the phase shift affects the waveform, but the type of dynamical regime (a comb in this case) is still the same for all the simulations. We have repeated this numerical experiment and verified the occurrence of this behaviour for other combinations of parameters, such as $L_\mathrm{ext,0}=20~$cm, $\epsilon^2=4~\%$ and $L_\mathrm{ext,0}=20~$cm, $\epsilon^2=1~\%$, where comb emission is obtained for all the considered values of external cavity length shift $\Delta L_\mathrm{ext}=n\lambda/4$, with n$=1,2,3,4$; for the combinations of$L_\mathrm{ext,0}=40~$cm, $\epsilon^2=4~\%$, $L_\mathrm{ext,0}=1~$m, $\epsilon^2=36~\%$, mixed state emission is obtained for all the considered values of external cavity length $\Delta L_\mathrm{ext}=n\lambda/4$. Therefore, for the case $I=1.08I_\mathrm{thr}$ (free-running CW as initial condition) these results show that in the short cavity regime an alternance between comb states and single mode emission occurs by performing a fine tuning of the external cavity on the $\lambda$ scale. On the contrary, if we consider the long cavity regime, a variation in the length of the external cavity on the wavelength scale does not alter the type of dynamic regime emitted by the laser, although it implies a slight variation in the obtained waveform.\\
\section{Estimation of the critical value of $\alpha$ factor for irregular dynamics}
\begin{figure}[t] 
\centering
\includegraphics[width=0.5\textwidth]{FiguresSupplementary/alphatuning.png}
\caption{OFC indicators $M_{\sigma_P}$ and $M_{\Delta\Phi}$ as a function of the $\alpha$ factor, for $L_\mathrm{ext}=1$~cm, $\epsilon^2=36~\%$, $\tau_\mathrm{e}=5$~ps, and other parameters as in Table I.}
\label{altun}
\end{figure}
\begin{figure} 
\centering
\includegraphics[width=0.5\textwidth]{FiguresSupplementary/irreg_con_BN.png}
\caption{Example of irregular regime for $\alpha=0.5$, $\tau_\mathrm{e}=5$~ps, $L_\mathrm{ext}=1$~cm, $\epsilon^2=36~\%$, and other parameters as in Table~I. (a) Temporal evolution of the output power. (b) Zoom on the first beatnote in the Power spectrum.}
\label{irreg2}
\end{figure}
We estimated the critical value of $\alpha$ for which irregular dynamics is observed, considering the case $L_\mathrm{ext}=1$~cm, $\epsilon^2=36~\%$, $\tau_\mathrm{e}=5$~ps, and simulating the QCL dynamics for the values of $\alpha$ between -0.1 and 0.7, with a step 0.1, and the other parameters fixed as in Table I. We chose this case because for $\alpha=-0.1$ we have a comb regime, while for $\alpha=0.7$ we report irregular dynamics, so that we are interested to estimate the value where the transitions between these two types of regimes occur. For a rigorous estimation, we calculate the comb indicators $M_{\sigma_P}$ and $M_{\Delta\Phi}$. In Fig.~\ref{altun} we show the plot of the OFC indicators for different values of $\alpha$ for the mentioned case, and we observe a clear transition from comb to unlocked dynamics for $\alpha=0.5$, a value which is compatible with THz QCLs.\cite{SilvestriReview, PiccardoReview} We verified that the unlocked states reported for $\alpha\geq0.5$ are actual irregular regimes and not mixed states (which are also unlocked according to the definition of the indicators, even if they present regular amplitude modulations with the periodicity given by the EC roundtrip). We remark that we referred to these states as "irregular" rather than "chaotic", because we did not assess their level of chaos using methods such as analyzing Lyapunov exponents, which would be necessary for accurately labeling them as "chaotic".\\
An example of these regimes, corresponding to $\alpha=0.5$, is reported in Fig.~\ref{irreg2}. It can be observed that the output power exhibits irregular modulations (Fig.~\ref{irreg2}(a)), which do no present the periodicity given by the roundtrip of the external cavity. For this reason, they can not be classified as mixed states. If we look at the zoom on the first beatnote in the power spectrum (Fig.~\ref{irreg2}(b)), we notice that this is split in several equidistant peaks spaced by the difference between FSR and $\mathrm{FSR_{ext}}$. In this case the nominal value of FSR is $20.8~$GHz, $\mathrm{FSR_{ext}}=15~$GHz, and the found spacing in Fig.~\ref{irreg2}(b) is $5.8~$GHz. This shows that these irregular states arise from the beating between the external and laser cavity modes. By looking at the maps of Figs.~4(c)-(d) in the manuscript, we notice that they occur in a region where the FSR and $\mathrm{FSR_{ext}}$ have the same order of magnitude. We understand, therefore, that in presence of high value of both $\alpha$ and feedback strength, if FSR and $\mathrm{FSR_{ext}}$ have the same order magnitude we do not obtain a comb emission but an unlocked dynamics as described.

\section{Fine Tuning of the external cavity length for a pre-existing frequency combs (OFC as initial condition)}\label{finetuningcomb}
\begin{figure} 
\centering
\includegraphics[width=0.42\textwidth]{FiguresSupplementary/Figura_7_mm_1_perc.png}
\caption{Fine tuning of external cavity length in the short cavity regime, starting from a free-running comb emission ($I=1.5I_\mathrm{thr}$). Temporal evolution of the output power (left) and power spectrum (right) for different values of the EC length shift $\Delta L_\mathrm{ext}$: (a) 0, (b) $\lambda/4$, (c) $\lambda/2$, (d) $3\lambda/4$, (e) $\lambda$. $\lambda$ is the central emission wavelength. $L_\mathrm{ext}=7~$mm, $\epsilon^2=1~\%$, $\alpha=$ -0.1; (b) $\tau_\mathrm{e}=$ 5~ps, other parameters as in Table~\ref{Table1}.}
\label{tuningcomb}
\end{figure}
We consider $I=1.5I_\mathrm{thr}$, corresponding to a free running comb emission. We add the feedback and, for some pairs ($L_\mathrm{ext}$, $\epsilon^2$) we observe still emission of an OFC, as depicted in the map of Fig.~5(a) in the main manuscript (see the light blue region). Let us consider a pair in the light blue region, i.e. $L_\mathrm{ext}=7~$mm, $\epsilon^2=1~\%$, which is also one of the cases displayed in Fig. 5(b) of the main manuscript. We perform a fine tuning of the EC length on the wavelength scale for this comb solution, reporting some differences with respect to the $I=1.08I_\mathrm{thr}$ case discussed in Sec.~\ref{finetuningCW}. In fact, we observe a passage from a comb regime (Fig.~\ref{tuningcomb}(a)) to irregular dynamics (Fig.~\ref{tuningcomb}(b)) when the EC length shift $\Delta L_\mathrm{ext}$ increases from 0 to $\lambda/4$, and then a return to comb for $\Delta L_\mathrm{ext}=\lambda/2$ (Fig.~\ref{tuningcomb}(c)). Then, we again obtain irregular dynamics for $\Delta L_\mathrm{ext}=3\lambda/4$, and a comb for $\Delta L_\mathrm{ext}=\lambda$. Therefore, our numerical simulations show that the comb solutions are obtained with periodicity $\lambda/2$.\\
If we consider a a case in the long cavity regime ($L_\mathrm{ext}=60~$cm, $\epsilon^2=1~\%$), we observe that the comb states alternate with mixed states, characterized by a multi-peaked beatnote, as shown in Fig.~\ref{tuningcomb2}
\begin{figure} 
\centering
\includegraphics[width=0.46\textwidth]{FiguresSupplementary/Figura_60_cm_1.png}
\caption{Fine tuning of external cavity length in the long cavity regime, starting from a free-running comb emission ($I=1.5I_\mathrm{thr}$). Temporal evolution of the output power (left) and power spectrum (right) for different values of the EC length shift $\Delta L_\mathrm{ext}$: (a) 0, (b) $\lambda/4$, (c) $\lambda/2$, (d) $3\lambda/4$, (e) $\lambda$. $\lambda$ is the central emission wavelength. $L_\mathrm{ext}=60~$cm, $\epsilon^2=1~\%$, $\alpha=$ -0.1; (b) $\tau_\mathrm{e}=$ 5~ps, other parameters as in Table~\ref{Table1}.}
\label{tuningcomb2}
\end{figure}
This scenario is the result of the nonlinear competition between the EC and laser cavity modes in the QCL medium. These results reproduce the experimental findings presented in ref.\cite{Liao22}, testifying the reliability of our simulator.

\section{Feedback regime map for a mid-IR QCL}
\begin{table}[h]
\begin{center}
\begin{tabular}{ |l|l| } 
\hline
 Homogeneous gain bandwidth & $\delta_\mathrm{hom}=1.1$~THz  \\
 Linewidth enhancement factor & $\alpha=0.7$   \\
 Central emission frequency & $\nu_0=30~$THz  \\
 Carrier lifetime & $\tau_\mathrm{e}=1~$ps  \\
 \hline
\end{tabular}
\caption{\label{Table2} Model parameters for the simulation of a mid-IR QCL.\cite{SilvestriReview,Silvestri20,PiccardoReview, Faist_2016,Faist_book}}
\end{center}
\end{table}
 In this section we replicate the study of Sec. III in the main manuscript for the case of a mid-IR QCL. We consider values of $\delta_\mathrm{hom}$, $\alpha$ factor, central emission frequency $\nu_0$, and carrier lifetime $\tau_\mathrm{e}$ that are typical of mid-IR QCLs, as indicated in Table~\ref{Table2}. For the ESMBE parameters not specified in Table~\ref{Table2} we utilize the values provided in Table~\ref{Table1}. The value bias current value is $I=1.08~I_\mathrm{thr}$, and corresponds to single mode emission in free-running operation, in analogy with the study presented in Fig.~2 in the main manuscript.\\
The feedback diagram is shown in Fig.~\ref{mappamidIR}.
\begin{figure} 
\centering
\includegraphics[width=0.5\textwidth]{FiguresSupplementary/MappaMidIR_new.png}
\caption{Feedback regime diagram for a mid-IR QCL. The values of $\epsilon$ and $L_\mathrm{ext}$ are chosen in the same intervals as in the feedback maps of Figs.~2 in the main manuscript. The bias current of the QCL is $I=1.08~I_\mathrm{thr}$, and corresponds to single mode emission in free-running operation.}
\label{mappamidIR}
\end{figure}
The higher value of gain bandwidth and $\alpha$ with respect to the THz QCL case of Fig.~2 leads to a larger number of modes involved in the competition inside the active medium. In fact, the phase-amplitude coupling provided by $\alpha$ favors the multimode dynamics of the laser,\cite{SilvestriReview,Opacak2019,Silvestri20} which can also manifest itself more easily in presence of a larger gain profile. For this reason, the dynamical scenario presented in the feedback map of Fig.~\ref{mappamidIR} is characterized by a large number of multimode states, both locked and unlocked. A comb region is observed for $4~\%<\epsilon^2<10~\%$ in the short cavity range ($L_\mathrm{ext}<1~$cm). However, for $\epsilon^2>10\%$ we report a majority of irregular regimes if $L_\mathrm{ext}<10~$cm. The high level of feedback, in combination with a larger $\alpha$ and more modes involved, implies an enhanced competition between external and laser cavity modes, and prevents the locking of the involved optical lines. Then, for $L_\mathrm{ext}>10~$cm we observe mixed states, in analogy with the THz QCL diagram of Fig.~2.\\
Furthermore, we notice that the parameter configuration discussed in this section differs from the case of Fig.~4(c) in the main manuscript for the larger value of $\delta_\mathrm{hom}$. This allows us to highlight the role of the gain bandwidth in the comb/multimode state formation in presence of an external target. In fact by comparing the two maps of Fig.~\ref{mappamidIR} and Fig.~4(c), we clearly observe that a larger gain bandwidth is crucial in order to have multimode regimes and comb formation in the short cavity range, and compensates the effect of the short carrier lifetime, which tends to keep stable the single-mode emission.\cite{mezzapesa2013,columbo2014}\\
\begin{figure} 
\centering
\includegraphics[width=0.4\textwidth]{FiguresSupplementary/Figura_comb_midIR.png}
\caption{Temporal evolution of the output power for feedback-induced HFCs in mid-IR QCLs, with harmonic order 5 (a), 6 (b) and 7 (c). They are obtained for the same feedback ratio $\epsilon^2=36~\%$ and for $L_\mathrm{ext}$ respectively of $3~$mm, $6~$mm, and $2~$mm. These combs are extracted from the map of Fig.~\ref{mappamidIR}.}
\label{combsmidIR}
\end{figure}
Finally, we observe that in the mid-IR QCL case, we report HFCs with order 5 (Fig.~\ref{combsmidIR}(a)), 6 (Fig.~\ref{combsmidIR}(b)), and 7 (Fig.~\ref{combsmidIR}(c)), that were not found with the parameters of a THz QCL (in that case the maximum reported order was 4). This shows that in correspondence of lower values of both polarization dephasing time $\tau_\mathrm{d}$ and carrier lifetime $\tau_\mathrm{e}$ (and therefore faster carrier and polarization dynamics), HFCs characterized by a faster time scale can be generated in presence of feedback.
\newpage
\bibliography{aipsamp}